\documentclass{aa}  
\usepackage[colorlinks=true,linkcolor=blue,citecolor    = blue]{hyperref}
\usepackage{graphicx}
\usepackage{txfonts}
\usepackage{float}
\usepackage{rotating,tabularx}
\usepackage{pdflscape}

\usepackage{multirow}
\usepackage{afterpage}
\usepackage{bigstrut}
\usepackage[table,xcdraw]{xcolor}
\usepackage[flushleft]{threeparttable }
\usepackage{amsmath}
\usepackage{enumitem}
\usepackage{amsfonts,color}
\usepackage{amssymb,float}
\usepackage[utf8]{inputenc}
\usepackage{color}
\usepackage{etoolbox}
\usepackage{appendix}
\usepackage{url}
\usepackage{dirtytalk}
\usepackage{float}
\usepackage{fontawesome}

\usepackage{booktabs}
\usepackage[graphicx]{realboxes}
\usepackage{adjustbox}
\usepackage{multirow}
\usepackage{rotating,tabularx}
\usepackage{afterpage}
\usepackage{tablefootnote}

\usepackage{tikz}

\usepackage{bigstrut}

\definecolor{lime}{HTML}{A6CE39}
\DeclareRobustCommand{\orcidicon}{%
    \begin{tikzpicture}
    \draw[lime, fill=lime] (0,0) 
    circle [radius=0.16] 
    node[white] {{\fontfamily{qag}\selectfont \tiny ID}};
    \draw[white, fill=white] (-0.0625,0.095) 
    circle [radius=0.007];
    \end{tikzpicture}
    \hspace{-2mm}
}

\foreach \x in {A, ..., Z}{%
    \expandafter\xdef\csname orcid\x\endcsname{\noexpand\href{https://orcid.org/\csname orcidauthor\x\endcsname}{\noexpand\orcidicon}}
}
\newcommand{\orcid}[1]{\href{https://orcid.org/#1}{\textcolor[HTML]{A6CE39}{\orcidicon}}}
\usepackage{graphicx}
\usepackage{txfonts}

\newcommand{\gaia}{\textit{Gaia}}

\newcommand{\baseline}{\texttt{baseline}}

\usepackage{arydshln}
\usepackage{dsfont}
\usepackage[ruled,vlined]{algorithm2e}
\SetKwComment{Comment}{/* }{ */}
\usepackage{makecell}

\begin{document} 
\title{A joint 1\% calibration of the RR Lyrae \& type-II Cepheid Leavitt laws yields homogeneous distances to 93 Galactic globular clusters}
\titlerunning{One percent calibration of the RRL and T2Cep Leavitt laws}

     \author{
   Bastian Lengen\inst{1}\orcid{0009-0007-8211-8262} \and Richard I. Anderson\inst{1}\orcid{0000-0001-8089-4419}  \and Mauricio Cruz Reyes \inst{1}\orcid{0000-0003-2443-173X}  \and Giordano Viviani \inst{1}\orcid{0009-0001-6201-2897}  }

   \institute{Institute of Physics, \'Ecole Polytechnique F\'ed\'erale de Lausanne (EPFL), Chemin Pegasi 51b, 1290 Versoix, Switzerland \\
   \email{bastian.lengen@epfl.ch, richard.anderson@epfl.ch}}

   \date{Received 10 September 2025} 

  \abstract{
  Recent work has established large samples of astrometrically confirmed RR Lyrae and type-II Cepheid members of Galactic globular clusters (GCs). Any given GC can contain multiple such stars at once, notably RR Lyrae stars pulsating in the fundamental mode (RRab) or the first overtone (RRc), and type-II Cepheids (T2Cep) of BL~Her and W~Vir types. Here, we present the first joint calibration of the Leavitt laws (LLs) exhibited by 802 RRab, 345 RRc, and 21 T2Cep stars anchored to trigonometric parallaxes. Using the third data release of the ESA {\it Gaia} mission (GDR3), we have calibrated the intercepts of the RRab and RRc Leavitt laws in the reddening-free {\it Gaia} Wesenheit magnitude, $M_G^W$, to better than $1.0\%$ in distance, and that of T2Cep to $1.3\%$, using a global fit to all data. The absolute scale is set by 37 nearby GCs with high-accuracy parallaxes while 56 additional GCs provide constraints on LL slopes as well as the LL intercept differences of RRc and T2Cep relative to RRab stars. Our global fit yields homogeneous high-accuracy distances of 93 GCs that show no evidence of bias for {\it Gaia} parallaxes of distant GCs. Control of systematics was demonstrated by 31 alternative fit variants, notably involving different treatments of metallicity effects, as well as by Markov Chain Monte Carlo analysis. Our results suggest that photometric metallicities of RR Lyrae stars require further improvements while also exhibiting possible signs of intra-cluster chemical inhomogeneity. This work lays the foundation for exploiting RRab, RRc, and T2Cep stars as high-accuracy standard candles for near-field cosmology and the extragalactic distance scale.}

   
   \maketitle

\section{Introduction} \label{sec:introduction}
Pulsating stars serve as crucial standard candles for the (extragalactic) distance scale thanks to the period–luminosity (PL) relations -- Leavitt laws \citep[LLs]{leavitt1912} -- they obey. The highest accuracy to date has been demonstrated for classical Cepheids, which can be observed in distant galaxies and have been extremely valuable for determining Hubble's constant  \citep[e.g.,][]{riess2022}. However, several types of population-II pulsating stars are available for distance determination in rather more nearby, old, and metal-poor populations. These include RR Lyrae (RRL) stars and type-II Cepheids (T2Cep), both of which are found in large numbers throughout the Milky Way halo, globular clusters (GCs), and nearby dwarf galaxies. Indeed, RRL were initially referred to as ``cluster-type variables'' \citep{smith2000}. 

RRL are horizontal branch stars that typically pulsate in either the fundamental mode (RRab) or the first overtone (RRc), with typical periods of $\sim0.6\,\mathrm{d}$ and $\sim0.3\,\mathrm{d}$, respectively. They are abundant: over 270,000 have been identified in the third data release of the ESA \gaia\ mission \citep[GDR3]{gaia2016, gaiaDR32023, clementini2023}, including in the LMC, SMC, M31, and numerous other Local Group systems \citep{clement2017, vazquez2023, cruzreyes2024}. RRL stars exhibit LLs in the infrared and in reddening-free magnitudes \citep[e.g.,][]{bhardwaj2017, neeley2019, braga2020}. 

T2Cep occupy the classical instability strip at luminosities exceeding the horizontal branch and are frequently grouped into three subclasses, by increasing period, including the BL~Her, W~Vir, and RV~Tau types. They are rarer but trace more advanced post-horizontal-branch evolutionary phases, including stars evolving toward or from the asymptotic giant branch \citep[for a review, cf.][]{bono2024}. In total, 1551 T2Cep have been identified in the \texttt{gaiadr3.vari\_cepheid} table \citep{ripepi2023}, and all T2Cep subtypes found in GCs exhibit a common LL \citep{cruzreyes2025}. At infrared wavelengths and Wesenheit magnitudes, T2Cep appear to extend the RRab LL to longer periods, which is particularly apparent in the Magellanic Clouds \citep{bhardwaj2017, soszynski2019}.

Several studies have presented calibrations of RRL and T2Cep LLs based on GDR3 parallaxes in conjunction with infrared photometry or using \gaia\ photometry to compute reddening-free Wesenheit magnitudes, $M_G^W$. For example, \citet{garofalo2022} used a hierarchical Bayesian method to calibrate the \gaia\ Wesenheit LL of field RRL, and \citet{bhardwaj2023} used GC distances from \citet{bailerjones2021} to calibrate RRL and T2Cep (separately) using near-infrared photometry. Both studies focused extensively on metallicity effects of RRL stars \citep{jurcsik1996}, which have been studied in a variety of photometric systems and using different approaches \citep[e.g.,][]{mullen2022, mullen2023, li2023, narloch2024}. While previous calibrations did include GDR3 parallax corrections provided by \citet{lindegren2021}, they did not correct for well-known residual parallax systematics demonstrated for stars brighter than $G \sim 12$\,mag by a variety of methods \citep[e.g.,][and references therein]{khan2023b}. 

Clusters hosting RRL or T2Cep offer compelling opportunities for improvements over the state-of-the-art. Firstly, their average parallaxes can be determined with high accuracy using stars in the magnitude range not affected by the residual bias, in analogy with classical Cepheids in open clusters \citep{riess2022gaia,cruzreyes2023}. Secondly, GCs hosting multiple specimens of a given type of pulsator provide valuable constraints on LL slopes. Thirdly, GCs hosting multiple pulsator types provide crucial constraints on the relative offsets between the RRab, RRc, and T2Cep LL intercepts. Fourthly, the second and third points also apply in the case when GC parallaxes are not determined with sufficient S/N for inversion, thus creating an opportunity for considering a much larger sample of GCs. Finally, the multiplicative effect on parallax induced by an additive offset in magnitude space \citep[cf.][]{Riess2021} allows to validate the accuracy of the GC parallaxes zero-point.

In this work, we have exploited these benefits by performing the first joint calibration of the LLs of RRab, RRc, and T2Cep stars across 93 GCs. Our methodology employs a global least-squares framework inspired by the SH0ES distance ladder \citep{riess2009, riess2022} that relies exclusively on GDR3 data, notably trigonometric parallaxes and photometry. With this approach, we sought to
\begin{itemize}
  \item calibrate the LLs of RRab, RRc, and T2Cep while accounting for their correlated zero points and treating metallicity dependencies consistently;
  \item derive accurate distances to all the GCs in the sample and to assess the validity of concerns raised about \gaia\ parallaxes in the low-S/N regime \citep{vasilievbaumgardt2021};
  \item provide a robust foundation for an independent Population-II distance ladder anchored directly to \gaia.
\end{itemize}

The structure of the paper is as follows. Section~\ref{sec:dataset} presents the data considered, including a description of the GCs and variable star samples (Sect.~\ref{sec:GCdata} and Sect.~\ref{sec:pulsating_star_sample}), photometry and reddening (Sect.~\ref{sec:photometry}), and metallicity measurements (Sect.~\ref{sec:metallicities}). Section~\ref{sec:methodolgy} describes the methodology, including the models (Sect.~\ref{sec:models}), fitting approach (Sect.~\ref{sec:fit}), MCMC framework (Sect.~\ref{sec:meth:MCMC}), outlier rejection (Sect.~\ref{sec:outliers}), and model variants (Sect.~\ref{sec:variants}). Section~\ref{sec:results} presents the main results, with baseline and variant fits. Section~\ref{sec:discussion} discusses the implications and comparison with the literature. Finally, Section~\ref{sec:conclusions} summarizes our findings.

\section{Data set} \label{sec:dataset}
This work exploits the astrometric and photometric capabilities of the \gaia\ mission \citep{gaia2016}, a space-based survey by the European Space Agency designed to map over a billion stars in the Milky Way with high-precision parallaxes, proper motions, and multi-band photometry. Specifically, we use data from \gaia\ Data Release 3 \citep{gaiaDR32023}, which provides the most precise all-sky measurements to date for large samples of variable stars.

Building on the results of \citet{cruzreyes2024, cruzreyes2025}, we incorporate average parallaxes for 170 GCs, allowing us to include RRL and T2Cep within these systems. This approach ensures a homogeneous, self-consistent dataset, enabling accurate luminosity and distance estimates while minimizing systematics.

\subsection{Galactic Globular Clusters}\label{sec:GCdata}
Our sample is taken from the 170 GCs listed in Table A.2 of \citet{cruzreyes2024}, which provides cluster-level parameters. In particular, it includes weighted mean parallaxes computed with the method of \citet{maizappelaniz2021}, incorporating the \citet{lindegren2021} parallax bias corrections and the covariance on small angular scales.  Among these, 115 host RRL, 39 contain T2Cep, and 37 include both types of variables, cf. Sect.\,\ref{sec:pulsating_star_sample}. These GCs offer a unique opportunity for calibrating distance scales, as their parallaxes enable direct geometric distance measurements, while their member standard candles provide a photometric approach. This combined approach makes GCs strong benchmarks for refining the distance ladder. They also allow for consistency checks between individual-star parallaxes and GC parallaxes, which should agree if no systematic biases are present.

\subsubsection{Magnitude-dependent parallax systematics} \label{sec:parallax systematics}
Even after applying the \citet{lindegren2021} correction to \gaia\ DR3 parallaxes, residual biases can remain. To assess their impact, we compare the parallaxes of individual cluster member stars to the weighted mean GC parallaxes from \citet{cruzreyes2024}. Previous studies report that cluster parallaxes generally show minimal or no residual offset, even in the low signal-to-noise (S/N) regime \citep{baumgardtvasiliev2021, appellaniz2022, riess2022gaia, cruzreyes2024}.

While a direct comparison between individual stars and cluster parallaxes is limited by the low S/N of single-star parallaxes, aggregating stars across similar magnitudes enables a more meaningful test. We therefore apply a rolling binning scheme in $G$-band magnitude, similar to a moving average, to boost S/N and reveal any systematic trends. This analysis includes all 427034 stars used in the GC parallax determinations.

\begin{figure}
    \centering
    \includegraphics{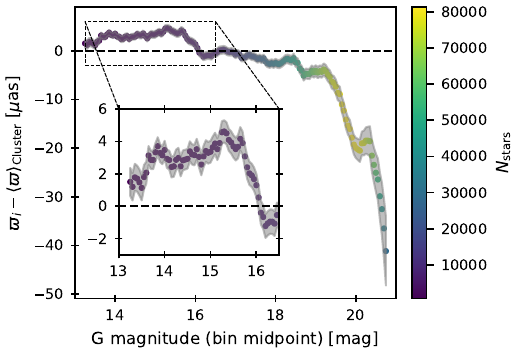}
    \caption{Mean residual parallax difference as a function of $G$-band magnitude, computed with rolling bins of width 0.5\,mag and step size 0.05\,mag. The $x$-axis marks the bin midpoints, while the $y$-axis gives the corresponding mean residual. Points are color-coded by the number of stars per bin, and the shaded area shows the $1\sigma$ uncertainty of the mean. The inset zooms into the range populated by most RRL and T2Cep stars, where a typical residual trend of $\sim$3–4\,$\mu$as is seen.} \label{fig:Gaia_varpi_Gmag}
\end{figure}

Figure~\ref{fig:Gaia_varpi_Gmag} shows the residual parallax offset as a function of \gaia\ $G$ magnitude. In the key range $G \in [13,16]$, where most RRL and T2Cep reside, we detect a mild systematic trend of $\sim$3–4\,$\mu$as. A sharper deviation appears near $G \approx 16$\,mag, coinciding with a change in \gaia's CCD window class configuration \citep{gaia2016}. However, this faint-end feature does not affect our analysis, as stars in this regime have low S/N and thus contribute little weight to the cluster parallax estimates.

\subsection{Pulsating star samples}\label{sec:pulsating_star_sample}
All the stars used in this study are members of GCs as identified by \citet{cruzreyes2024, cruzreyes2025}, and are drawn from the \gaia\ DR3 catalog. The RRL sample is listed in Table 1 of \citet{cruzreyes2024}, and the T2Cep sample in Table A.1 of \citet{cruzreyes2025}. We retain only stars flagged as part of the final samples (\texttt{Final=True}) and with valid (non-NaN) values in the \texttt{Posterior} column, ensuring the exclusion of stars near GC centers where astrometric and photometric measurements are typically less reliable.

We further restrict the sample to stars with Specific Object Studies (SOS) photometry (\texttt{SOS=True}), a dedicated pipeline optimized for pulsating variables \citep{ripepi2019, gaiaDR32023, clementini2023}. Additional photometric quality cuts, detailed in Table~\ref{tab:photometric_cuts}, are applied. We also retain only stars for which the \citet{lindegren2021} parallax correction is defined, and exclude stars with missing (NaN) values in parameters relevant to the analysis.

\begin{table}
    \centering
    \caption{Photometric cuts and anchors criteria applied to the RRL and T2Cep samples.}
    \label{tab:photometric_cuts}
    \begin{tabular}{c c}
        \hline 
        \textbf{Photometric cuts} & \quad\textbf{Anchors criteria}         \\ 
        Individual star & \quad Cluster \\ \hline \hline
        \texttt{ipd\_frac\_multi\_peak} \textless 10& \quad$\varpi/\sigma_\varpi > 10$        \\ 
        \texttt{num\_clean\_epochs\_g} \textgreater 10 & \quad$E(B-V) < 1.0$ \\
        \texttt{num\_clean\_epochs\_bp} \textgreater 10 & \\
        \texttt{num\_clean\_epochs\_rp} \textgreater 10 & \\ \hline
    \end{tabular}
\end{table}

\subsection{Photometry and Wesenheit magnitudes} \label{sec:photometry}
It is well established that RRL do not follow a PLR in visual bands such as \( V \), consistent with their characterization as horizontal branch stars \citep{catelan2004, bhardwaj2022}. However, they follow a PLR at longer wavelengths or in reddening-free bands such as optical Wesenheit magnitudes, where the inclusion of color terms restores a tight PL correlation \citep{marconi2015, braga2015, neeley2017}. 

In this work, we use Wesenheit magnitudes as defined in the \gaia\ photometric system \citep{madore1982, ripepi2019}. Specifically, we adopt the optical Wesenheit magnitude \( M_G^W \), constructed from the \textit{G}, \textit{BP}, and \textit{RP} bands:
\begin{equation}\label{eq:Wesenheit}
    m^W_G = m_G - R^W_G \left(m_{BP}-m_{RP}\right)
\end{equation}
Here, \( m_G \), \( m_{BP} \), and \( m_{RP} \) are the observed magnitudes in the respective bands, and the reddening coefficient \( R^W_G \) is defined as
\begin{equation}\label{eq:Reddening}
    R^W_G = \frac{A_G}{A_{BP}-A_{RP}}
\end{equation}
We adopt a fixed value of \( R^W_G = 1.867 \) for both RRL and T2Cep, following \citet{cruzreyes2024}.

\subsubsection{Uncertainties reassessment} 
Photometric uncertainties increase with magnitude, reflecting lower S/N for fainter stars. However, a small subset of stars in the \gaia\ SOS catalog report unrealistically small uncertainties compared to others of similar brightness \citep{cruzreyes2024}. For example, the star with \texttt{source\_id} = 4295843576778048384 has a $BP$-band uncertainty ($\texttt{int\_average\_bp\_error}$) of just $7 \times 10^{-15}$\,mag, orders of magnitude below typical values for that magnitude range. Since the Wesenheit magnitude propagates errors from multiple bands, an underestimated uncertainty in a single band can artificially reduce the total error, giving the star disproportionate weight in the fit.

To address this, we examine the typical photometric uncertainties as a function of magnitude and identify outliers. We then reassess each star’s reported uncertainty by comparing it to the median uncertainty of stars in the same 1\,mag-wide bin. If a star’s uncertainty in any band is more than three standard deviations below the bin median, we replace it with the bin’s median value. Stars with typical or larger-than-expected uncertainties remain unchanged. This correction affects approximately 10\% of stars, depending on filter and type of variable, and ensures that stars of similar brightness contribute comparably to the model fit.

In addition, PLRs exhibit intrinsic scatter due to the finite width of the instability strip, which introduces luminosity dispersion at fixed period \citep{marconi2015, neeley2017}. To account for this, we include an intrinsic scatter term following the SH0ES approach for Cepheids \citep{riess2022}. Specifically, we add 0.045\,mag in quadrature to the Wesenheit magnitude uncertainty for both RRL and T2Cep. This value is smaller than the intrinsic scatter typically observed for classical Cepheids and is consistent with the dispersion seen in our baseline fit (see Sect.~\ref{sec:LL_residuals}).

\subsection{Metallicities} \label{sec:metallicities}
Metallicity affects the luminosity of pulsating stars and is a known source of scatter in PLR. Its influence on PL-based distances is widely discussed, particularly in the context of classical Cepheids \citep{romaniello2008, breuval2021, riess2022}. Properly accounting for metallicity is therefore essential to avoid biases in distance calibrations, especially when comparing stars across a wide range of environments.

For T2Cep, we adopt the metallicity of their host cluster, as listed in the \citet{harris2010} catalog. These values are based on spectroscopic measurements of red giant branch stars and serve as fiducial [Fe/H] estimates for Milky Way GCs. This approach assumes chemical homogeneity within clusters and does not account for intra-cluster variation.

The same catalog values can also be used for RRL. However, RRL light curves correlate with metallicity \citep{jurcsik1996}, enabling photometric [Fe/H] estimates derived from their pulsation morphology. We adopt such light-curve-based proxies as relative indicators of chemical composition. While the absolute accuracy of these calibrations may be uncertain, this has limited impact on our analysis: PL-based distances are primarily sensitive to relative metallicity differences. Since the calibration is applied homogeneously across our sample, it still allows robust measurement of the PLR’s metallicity dependence without introducing systematic distance offsets.

For \gaia\ G-band SOS light curves \citep{gaiaDR32023, clementini2023}, we adopt the recent calibrations by \citet{li2023}, derived from a sample of over 2700 RRL with spectroscopic metallicities. The $P$–$\phi_{31}$–$R_{21}$–[Fe/H] relation is used for RRab stars, and the $P$–$R_{21}$–[Fe/H] relation for RRc stars, as defined below:
\begin{align}
    \begin{split}
        [\textrm{Fe/H}]_{RRab} &= -1.888 - 5.772\cdot(P - 0.6) \\
        & \quad + 1.065\cdot(R_{21} - 0.45) + 1.090\cdot(\phi_{31} - 2) 
    \end{split} \label{RRab_Li}\\
    \begin{split}
        [\textrm{Fe/H}]_{RRc} &= -1.737-9.968\cdot(P-0.3) \\
        & \quad  -5.041\cdot(R_{21}-0.20)
    \end{split} \label{RRc_Li}
\end{align}

To assess the impact of systematics, we also consider an alternative calibration by \citet{muraveva2025}. Although the functional form differs from that of \citet{li2023}, it is similarly based on the idea that the shape of RRL light curves encodes information about metallicity. Specifically, this calibration employs different combinations of light-curve parameters:
\begin{align}
    [\textrm{Fe/H}]_{RRab} &= -0.37 -5.55\cdot P +0.94\cdot \phi_{31} \label{RRab_Muraveva}\\
    [\textrm{Fe/H}]_{RRc} &= 0.80 -8.54\cdot P +0.23\cdot \phi_{31}-9.34\cdot A_2\label{RRc_Muraveva}
\end{align}
These calibrations are particularly advantageous as they are derived from a large homogeneous dataset using \gaia\ photometry, allowing us to estimate metallicities homogeneously for the RRL in our sample, many of which lack direct spectroscopic measurements. This approach enables us to account for both inter- and intra-cluster metallicity variations while minimizing systematic biases. 

\section{Methodology} \label{sec:methodolgy}
Our method simultaneously calibrates the PLR for RRL and T2Cep by leveraging the precise average parallaxes of GCs. The objective is to jointly fit the various PLRs (RRab, RRc, T2Cep) across different clusters. Any cluster containing multiple stars of a given type (e.g., RRab) contributes information on the slope of the corresponding PLR, regardless of its distance precision. Moreover, GCs hosting multiple types of pulsating stars provide valuable constraints on the relative luminosity scales between the different candles. To ensure consistency and precision in the calibration, we adopt the methodology employed in the Cepheid distance ladder, following the approach developed by the SH0ES team \citep{riess2016, riess2022}.

This method requires working in magnitude space, which involves inverting parallaxes, a process that can introduce biases when parallax uncertainties are large. To address this, we classify clusters into two categories based on the quality of their parallaxes. Clusters with reliable parallaxes (typically nearer, with low extinction and high S/N) are designated as anchors, as they provide the geometric foundation for anchoring the LLs. More distant clusters with less reliable parallaxes or significant extinction are designated as hosts. Their distances are not directly constrained by their parallaxes but are instead inferred from the fitted PLR, allowing us to bypass the use of their less reliable parallax measurements.

This two-class classification enables us to incorporate variable stars from a wide range of GCs. Even when their geometric distances are uncertain, these clusters still contribute essential information, particularly in constraining the slopes of the PLR. The criteria used to classify clusters as anchors or hosts are summarized in Table~\ref{tab:photometric_cuts}.

\subsection{Models} \label{sec:models}
For each type of variable, a separate PLR is fitted. The model differs slightly depending on whether the star belongs to a host or an anchor cluster. For stars in hosts, where the distance modulus $\mu_{0,i}$ is a free parameter, the PLR for RRab stars is
\begin{align}
    m^W_{i,j} = \mu_{0,i} 
              &+ \alpha_{0,\mathrm{RRab}} \notag \\
              &+ \beta_{0,\mathrm{RRab}}\left(\log P_{i,j} - \log P_{0,\mathrm{RRab}}\right) \notag \\
              &+ \alpha_{M,\mathrm{RRab}}\left(\left[\textrm{Fe/H}\right]_{i,j} - \left[\textrm{Fe/H}\right]_{0,\mathrm{RRab}}\right) \label{eq:host_RRab}
\end{align}
here, indices $i$ and $j$ refer to the cluster and the individual star within that cluster, respectively. The quantity $m^W_{i,j}$ denotes the observed Wesenheit magnitude of star $j$ in cluster $i$, and $\mu_{0,i}$ is the extinction-corrected distance modulus of that cluster. The term $\alpha_{0,\mathrm{RRab}}$ represents the absolute Wesenheit magnitude at the pivot period $P_{0,\mathrm{RRab}}$ and metallicity $[\textrm{Fe/H}]_{0,\mathrm{RRab}}$, while $\beta_{0,\mathrm{RRab}}$ and $\alpha_{M,\mathrm{RRab}}$ are the slopes with respect to $\log P$ and $[\textrm{Fe/H}]$.

For RRc and T2Cep, we adopt the same model structure but calibrate their absolute magnitudes relative to RRab:
\begin{align}
    m^W_{i,j} = \mu_{0,i} 
              &+ \alpha_{0,\mathrm{RRab}} + \Delta\alpha_{0,\mathrm{RRc/T2Cep}} \notag \\
              &+ \beta_{0,\mathrm{RRc/T2Cep}}\left(\log P_{i,j} - \log P_{0,\mathrm{RRc/T2Cep}}\right) \notag \\
              &+ \alpha_{M,\mathrm{RRc/T2Cep}}\left(\left[\textrm{Fe/H}\right]_{i,j} - \left[\textrm{Fe/H}\right]_{0,\mathrm{RRc/T2Cep}}\right) \label{eq:host_RRc/T2Cep}
\end{align}
The relative offset $\Delta\alpha_{0,\mathrm{RRc}}$ or $\Delta\alpha_{0,\mathrm{T2Cep}}$ quantifies the difference in absolute magnitude, at fixed period and metallicity, between RRab and the other variable types. This formulation enables a consistent calibration across all pulsator classes while anchoring the absolute scale to RRab stars.

With this structure, the global fit is anchored by a single absolute zero point ($\alpha_{0,\mathrm{RRab}}$), while relative offsets ($\Delta\alpha_{0,\mathrm{RRc/T2Cep}}$) capture the luminosity differences between classes. If each class were instead fitted with its own independent zero point, all three calibrations would shift coherently with any change in the anchor distances, leading to strong correlations between the zero points and the GC distance moduli. Using one common $\alpha_0$ absorbs this global correlation, while the relative offsets are sensitive only to differences between classes and remain largely decoupled from the distance scale. In practice, this parametrization improves consistency across classes and confines the dominant uncertainty to $\alpha_{0,\mathrm{RRab}}$, while $\Delta\alpha_0$ is constrained primarily by clusters that host multiple types of variables.

We adopt pivot values $P_0$ and $[\textrm{Fe/H}]_0$ near the median period and metallicity of each class to minimize covariances between fit parameters and to improve numerical stability. These values are chosen close to the sample averages and rounded for convenience, facilitating the interpretation of the fitted slopes and intercepts.

For anchors, the equations take a slightly different form: the geometric distance modulus appears on the left-hand side (and is therefore not a fitted parameter), while a cluster-specific correction term $\Delta\mu_{0,i}$ is introduced on the right-hand side to account for residuals relative to the geometric distance:
\begin{align}
    m^W_{i,j} - \mu_{0,i} = \Delta\mu_{0,i} 
              &+ \alpha_{0,\mathrm{RRab}} \notag \\
              &+ \beta_{0,\mathrm{RRab}}\left(\log P_{i,j} - \log P_{0,\mathrm{RRab}}\right) \notag \\
              &+ \alpha_{M,\mathrm{RRab}}\left(\left[\textrm{Fe/H}\right]_{i,j} - \left[\textrm{Fe/H}\right]_{0,\mathrm{RRab}}\right) \label{eq:anchor_RRab}
\end{align}
\begin{align}
    m^W_{i,j} - \mu_{0,i} = \Delta\mu_{0,i} 
              &+ \alpha_{0,\mathrm{RRab}} + \Delta\alpha_{0,\mathrm{RRc/T2Cep}} \notag \\
              &+ \beta_{0,\mathrm{RRc/T2Cep}}\left(\log P_{i,j} - \log P_{0,\mathrm{RRc/T2Cep}}\right) \notag \\
              &+ \alpha_{M,\mathrm{RRc/T2Cep}}\left(\left[\textrm{Fe/H}\right]_{i,j} - \left[\textrm{Fe/H}\right]_{0,\mathrm{RRc/T2Cep}}\right) \label{eq:anchor_RRc/T2Cep}
\end{align}
The additional term $\Delta\mu_{0,i}$ captures small residual offsets between the model-predicted and geometric distance moduli. This term is constrained by
\begin{equation}
    0 = \Delta\mu_{0,i} \pm \sigma\left(\mu_{0,i}\right) \label{eq:constraint}
\end{equation}
where $\sigma\left(\mu_{0,i}\right)$ is the uncertainty on the parallax-based distance modulus. This treatment, following \citet{riess2022}, allows the use of multiple GC anchors to geometrically calibrate the global fit.

The equations are formulated in magnitude space, whereas \gaia\ provides parallaxes. This requires careful treatment when converting between distances and magnitudes. Estimating distances via $d = 1/\varpi$ becomes increasingly biased when the relative parallax uncertainty exceeds 0.1 \citep{luri2018}. To mitigate this, we apply an S/N cut (see Table~\ref{tab:photometric_cuts}) when selecting anchors, ensuring that only parallaxes with sufficient precision are used. This allows us to safely invert the parallaxes without introducing significant bias. The corresponding transformation to distance modulus is
\begin{equation}
    \mu = 5\log\left(d\right)-5 = 5\log\left(\frac{1}{\varpi}\right)-5
\end{equation}
where $\mu$ is the distance modulus in [mag], $d$ the distance in [pc], and $\varpi$ the parallax in [arcsec].

\subsection{Joint fit} \label{sec:fit}
The full definitions of the matrices and the explicit derivation of the least-squares solution are provided in Appendix~\ref{app:mathematical_details}. In short, the fit reduces to solving a system of linear equations, where each star and each geometric constraint contributes one relation. These can be expressed in matrix form as
\begin{equation} \label{eq:y=Lq}
    \boldsymbol{y} = \boldsymbol{L} \cdot \boldsymbol{q},
\end{equation}
with $\boldsymbol{y}$ denoting the left-hand side of Eqs.~\eqref{eq:host_RRab}--\eqref{eq:constraint}, $\boldsymbol{L}$ the design matrix containing the known coefficients from the right-hand side, and $\boldsymbol{q}$ the vector of parameters to be estimated. In the baseline case, $\boldsymbol{L}$ is a large matrix ($1207 \times 102$), but its block structure allows for an intuitive representation of the problem.

\subsection{Markov Chain Monte Carlo\label{sec:meth:MCMC}}
The Markov Chain Monte Carlo (MCMC) approach extends the least-squares solution by exploring the full posterior distribution of the parameters. This provides not only central values but also a comprehensive assessment of the associated uncertainties. Specifically, we infer the posterior distribution $p(\mathbf{q} \mid \mathbf{L}, \mathbf{C}, \mathbf{y})$ of the model parameters $\mathbf{q}$ given the observational data $\mathbf{y}$, the design matrix $\mathbf{L}$, and the associated uncertainties in $\mathbf{C}$:
\begin{equation} \label{eq:prior_distribution_general}
    p(\mathbf{q} \mid \mathbf{L}, \mathbf{C}, \mathbf{y}) \propto p(\mathbf{q})~ p(\mathbf{y} \mid \mathbf{L}, \mathbf{C}, \mathbf{q})
\end{equation}
Here $p(\mathbf{q})$ denotes the prior distribution, and $p(\mathbf{y} \mid \mathbf{L}, \mathbf{C}, \mathbf{q})$ is the likelihood. Assuming a Gaussian likelihood with diagonal covariance $\mathbf{C} = \mathrm{diag}(\sigma_1^2, \dots, \sigma_n^2)$, we write
\begin{equation} \label{eq:full_likelihood}
    p(\mathbf{y} \mid \mathbf{L}, \mathbf{C}, \mathbf{q}) = \prod_i\frac{1}{\sqrt{2\pi}\sigma_i}\exp\left(-\frac{(y_i - y_{\text{fit}, i})^2}{2\sigma_i^2}\right)
\end{equation}
which leads to a log-likelihood of the form
\begin{equation} \label{eq:prior_distribution_with_likelihood}
    \ln p(\mathbf{y} \mid \mathbf{L}, \mathbf{C}, \mathbf{q}) \propto -\frac{1}{2} \sum_i \chi^2_i
\end{equation}
with
\begin{equation} \label{eq:chi2_cholesky}
    \chi^2 = (\mathbf{y} - \mathbf{Lq})^T ~ \tilde{\mathbf{C}}_L^{-1} ~ (\mathbf{y} - \mathbf{Lq})
\end{equation}
where $\tilde{\mathbf{C}}_L$ is the Cholesky decomposition of the diagonal covariance matrix $\mathbf{C}$. Although $\mathbf{C}^{-1}$ is analytically straightforward, the Cholesky decomposition improves numerical stability, particularly for large or ill-conditioned systems.  

For the prior $p(\mathbf{q})$, we adopt a uniform distribution centered on the analytic least-squares solution $\boldsymbol{\mu}$ from Eq.~\eqref{eq:q_fit}:
\begin{equation}
    q_i \sim \mathrm{Uniform}(\mu_i - 10\sigma_i, \mu_i + 10\sigma_i)
\end{equation}
a broad prior (20$\sigma$ width) that allows the chains to explore the parameter space freely while remaining informed by the analytic solution.

Sampling is performed with the \texttt{emcee} Python package \citep{foreman-mackey2013emcee}, using 500 walkers initialized uniformly within the prior bounds. Convergence is monitored via the autocorrelation time $\tau$, estimated following \citet{goodman2010}. A burn-in phase of $5\tau$ is discarded, after which the chains are extended to a total length of $\sim100\tau$, ensuring adequate sampling of the posterior. The final posteriors lie well within the central region of the prior range, confirming that the prior bounds do not bias the inferred parameters.

\subsection{Outlier rejection} \label{sec:outliers} 
To mitigate the impact of potential outliers that pass through the selection criteria, an outlier-rejection algorithm is implemented within the fitting routine. The procedure follows the standard $\sigma$-clipping approach, in which the most deviant point is removed at each iteration until no star deviates by more than $\kappa\sigma$ from the fit \citep[e.g.,][]{kodric2026}.  

Because the dataset contains multiple types of variable stars with distinct intrinsic scatter, the classical single-population $\sigma$-clipping method is not directly applicable. Both the dispersion $\sigma$ and the rejection threshold $\kappa$ differ between stellar populations, making a uniform rejection criterion unsuitable. To account for this, a modified algorithm (see Appendix~\ref{app:outlier_algorithm}, Alg.~\ref{alg:kappa}) is adopted that incorporates population-dependent scatter. At each iteration, the most deviant star relative to the scatter of its own population is removed, and the process continues until the stopping criterion for each population is satisfied.

The values of $\kappa_i$ for different populations can be freely chosen. Unless otherwise stated, they are set using Chauvenet's criterion for each population.

\subsection{Baseline fit and variants} \label{sec:variants}
To assess the robustness of the results, a \baseline\ model is defined that incorporates the full dataset and reflects the most comprehensive set of assumptions. This model serves as the primary reference throughout the analysis. Its configuration is summarized in Table~\ref{tab:baseline_configuration}.

To evaluate the sensitivity of the conclusions to specific modeling choices and potential sources of systematic error, a series of targeted variants is constructed. Each variant isolates the impact of a particular assumption, input, or methodological choice. The variants are grouped into categories, each testing different sources of systematics in the analysis.
\begin{enumerate} 
    \item \textbf{Baseline}: The primary reference model.
    \item \textbf{Outlier rejection}: Sensitivity of the baseline model to different outlier-rejection criteria.
    \begin{itemize}
        \item[\textbullet] \underline{No OR}: No outlier rejection.
        \item[\textbullet] \underline{$\kappa=5.0$}: Weak rejection with $\kappa=5.0$.
        \item[\textbullet] \underline{$\kappa=3.5$}: Same as above, with $\kappa=3.5$.
        \item[\textbullet] \underline{$\kappa=3.0$}: Same as above, with $\kappa=3.0$.
        \item[\textbullet] \underline{$\kappa=2.5$}: Strict rejection with $\kappa=2.5$.
    \end{itemize}
    \item \textbf{Metallicities}: Impact of different metallicity calibrations.
    \begin{itemize}
        \item[\textbullet] \underline{Muraveva}: Calibration from \citet{muraveva2025} for RRab and RRc.
        \item[\textbullet] \underline{Cluster}: GC metallicities from \citet{harris2010} for all candles.
    \end{itemize}
    \item \textbf{Parallaxes}: Alternative determinations of GC parallaxes.
    \begin{itemize}
        \item[\textbullet] \underline{Median}: Median parallaxes instead of weighted mean.
        \item[\textbullet] \underline{$G<19$}: Weighted mean parallaxes using only stars with $G<19$ mag.
        \item[\textbullet] \underline{$14.5<G<19$}: Weighted mean parallaxes using only stars with $14.5<G<19$.
        \item[\textbullet] \underline{VB21}: Parallaxes from Table A.1 of \citet{vasilievbaumgardt2021} (VB21).
    \end{itemize}
    \item \textbf{Models}: Alternative assumptions to Eqs.~\eqref{eq:host_RRab}--\eqref{eq:anchor_RRc/T2Cep}.
    \begin{itemize}
        \item[\textbullet] \underline{No RRab}: Excluding RRab.
        \item[\textbullet] \underline{No RRc}: Excluding RRc.
        \item[\textbullet] \underline{No T2Cep}: Excluding T2Cep.
        \item[\textbullet] \underline{$\beta_{0,\textrm{RRc}}=\beta_{0,\textrm{RRab}}$}: Common slope $\beta$ for RRab and RRc. 
        \item[\textbullet] \underline{$\beta_{0,\textrm{T2Cep}}=\beta_{0,\textrm{RRab}}$}: Common slope $\beta$ for RRab and T2Cep.
        \item[\textbullet] \underline{$\beta_{0,\textrm{T2Cep}}=\beta_{0,\textrm{RRc}}$}: Common slope $\beta$ for RRc and T2Cep.
        \item[\textbullet] \underline{$\beta_{0,\textrm{T2Cep}}=\beta_{0,\textrm{RRc}}=\beta_{0,\textrm{RRab}}$}: All types share the same slope $\beta$.
        \item[\textbullet] \underline{$\alpha_{M,\textrm{RRc}}=\alpha_{M,\textrm{RRab}}$}: Common metallicity dependence $\alpha_M$ for RRab and RRc.
        \item[\textbullet] \underline{No $\alpha_M$}: Metallicity dependence ignored ($\alpha_M=0$).
        \item[\textbullet] \underline{$\beta_0\rightarrow\beta_0 + \beta_M[\textrm{Fe/H}]$}: Period slope $\beta$ depends on metallicity.
        \item[\textbullet] \underline{$\beta_0\rightarrow\beta_0 + \beta_M[\textrm{Fe/H}]+$ No $\alpha_M$}: Same as above, but no metallicity effect on the intercept.
        \item[\textbullet] \underline{$\beta\rightarrow\beta_0 + \beta_M[\textrm{Fe/H}]_{Cl}$}: Same as above, with cluster metallicities from \citet{harris2010}.
        \item[\textbullet] \underline{$\beta\rightarrow\beta_0 + \beta_M[\textrm{Fe/H}]_{Cl}+$ No $\alpha_M$}: Same as above, but no metallicity effect on the intercept, using cluster metallicities.
    \end{itemize}
    \item \textbf{Pre-processing}: Sensitivity to pre-processing parameters.
    \begin{itemize}
        \item[\textbullet] \underline{$\sigma_{\mathrm{extra}}=0$}: No additional scatter from the instability strip.
        \item[\textbullet] \underline{$\sigma_{\mathrm{extra}}=0.025$}: Extra scatter decreased by 50\% relative to \baseline.
        \item[\textbullet] \underline{$\sigma_{\mathrm{extra}}=0.075$}: Extra scatter increased by 50\% relative to \baseline.
    \end{itemize}
    \item \textbf{Anchor criteria}: Alternative thresholds for identifying GC anchors.
    \begin{itemize}
        \item[\textbullet] \underline{$S/N=5$}: Threshold set to half the \baseline\ value.
        \item[\textbullet] \underline{$S/N=20$}: Threshold set to twice the \baseline\ value.
        \item[\textbullet] \underline{$E(B-V)=\infty$}: No extinction-based criterion.
        \item[\textbullet] \underline{$E(B-V)=0.5$}: Threshold set to half the \baseline\ value.
    \end{itemize}
\end{enumerate}

\begin{table}
\centering
\renewcommand{\arraystretch}{1.3}
\caption{Overview of the baseline fit configuration.} \label{tab:baseline_configuration}
\resizebox{\columnwidth}{!}{
\begin{tabular}{|lcccclc|}
\hline
\textbf{Variables} & $\mathbf{\boldsymbol{\log} P_0}$ & $\mathbf{[Fe/H]_0}$ & $\mathbf{[Fe/H]}$ & $\boldsymbol{\kappa}$ & $\boldsymbol{\sigma_{\mathrm{extra}}}$ & $\boldsymbol{R^W_G}$ \\
\hline
\textbf{RRab} & $-0.25$ & $-1.5$ & Phot.\tablefootmark{1}& Chauvenet & $0.045$ & $1.867$ \\
\textbf{RRc}  & $-0.5$  & $-1.5$ & Phot.\tablefootmark{1} & Chauvenet & $0.045$ & $1.867$ \\
\textbf{T2Cep} & $1$     & $-1.5$ & GC\tablefootmark{2}  & Chauvenet & $0.045$ & $1.867$ \\
\hline
\multicolumn{3}{|l}{\textbf{GC Anchors Selection:}} & \multicolumn{1}{l}{S/N $\geq 10$} & \multicolumn{3}{l|}{E(B$-$$V$) $\leq 1$} \\
\hline
\end{tabular}}
\tablefoot{
\tablefootmark{1} Photometric metallicities from \citet{li2023}. 
\tablefootmark{2} GC metallicities from \citet{harris2010}.
}
\end{table}

\section{Results} \label{sec:results}
This section presents the LL calibrations from the \baseline\ analysis (Sect.~\ref{sec:results_baseline}), with uncertainties characterized using the MCMC approach (Sect.~\ref{sec:MCMC}). The corresponding distance estimates to GCs are given in Sect.~\ref{sec:GC_distances}, while Sect.~\ref{sec:result_variants} examines the impact of model variants. The complete set of best-fit parameters, including both the \baseline\ and all variant models, is listed in Table~\ref{tab:fit_results}. The distances to anchor and host GCs for the \baseline\ configuration are reported in Tables~\ref{tab:result_anchor} and \ref{tab:result_host}.

\subsection{Baseline results} \label{sec:results_baseline}
\begin{figure*}
    \centering
    \includegraphics{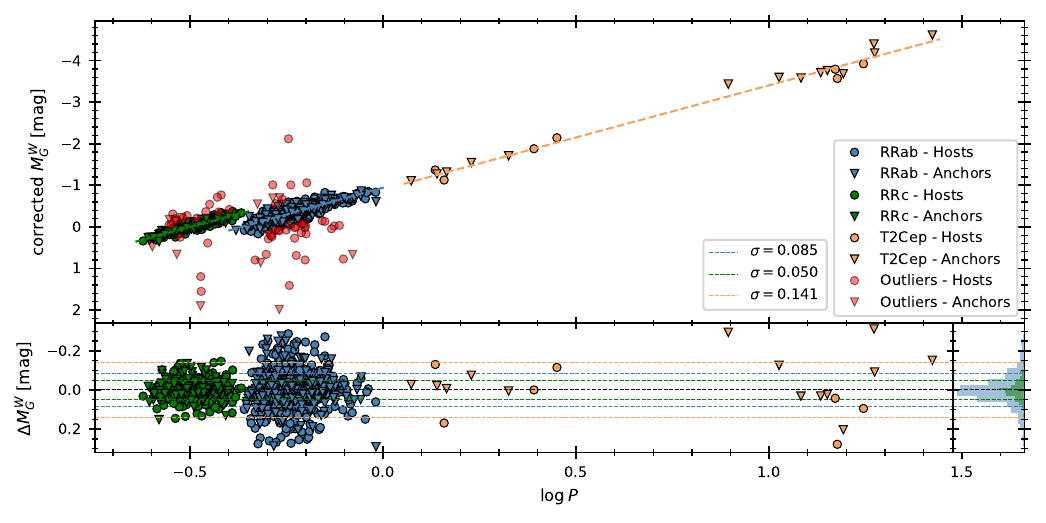}
    \caption{Baseline joint fit to fundamental-mode, first-overtone RR Lyrae, and type-II Cepheids in Galactic GC. Colors indicate different stellar types: blue for RRab, green for RRc, yellow for T2Cep, and red for stars excluded by the outlier rejection. Circles mark stars in hosts, while downward-pointing triangles denote stars in anchors. \textbf{Top:} Absolute magnitudes, corrected for metallicity, plotted against $\log P$. For hosts, $M^W_{i,j} = m^W_{i,j} - \mu_{0,i} - \alpha_M([\textrm{Fe/H}]_{i,j} - [\textrm{Fe/H}]_0)$; for anchors, an additional $-\Delta\mu_{0,i}$ term applies. Colored dashed lines show the fitted PLR. \textbf{Bottom:} Residuals between the corrected magnitudes and the fitted PLR, with dashed lines marking the residual standard deviation. The inset (bottom right) shows the residual distributions.}
    \label{fig:baseline_Global_LL}
\end{figure*}

Figure~\ref{fig:baseline_Global_LL} presents the data and fit from the \baseline\ analysis and illustrates the residual scatter for RRab, RRc, and T2Cep. The absolute magnitudes of all stars are standardized for metallicity differences to properly reflect the scatter in the residuals. The three types of variables occupy distinct period ranges with minimal overlap. The bottom panel shows the residuals, where RRab and T2Cep exhibit larger scatter ($\sigma = 0.085$\,mag and $\sigma = 0.141$\,mag, respectively) compared to RRc ($\sigma = 0.050$\,mag). The fit yields $\chi^2 \approx 1089$ for 1101 degrees of freedom, indicating good agreement between model and data. During the fitting process, Chauvenet’s criterion results in threshold values of $\kappa = 3.44$ for RRab, $3.20$ for RRc, and $2.26$ for T2Cep. Based on these thresholds, the outlier-rejection algorithm excludes about 5\% of the total sample: 43 RRab, 21 RRc, and no T2Cep.

Equations~\eqref{eq:result_RRab}--\eqref{eq:result_T2Cep} provide the \gaia-based Wesenheit absolute-magnitude calibrations derived from the \baseline\ fit for RRab, RRc, and T2Cep. The fiducial magnitudes for RRc and T2Cep are computed from the fitted $\Delta\alpha_0$ values (cf.~Eq.~\eqref{eq:anchor_RRab}), with uncertainties added in quadrature (cf.~Table~\ref{tab:fit_results}):
\begin{align}
M_{\mathrm{RRab}} &= (-0.300 \pm 0.020) \notag \\
&+ (-2.582 \pm 0.047)\cdot\left(\log P + 0.25\right) \notag \\
&+ (-0.028 \pm 0.008)\cdot\left([\mathrm{Fe/H}] + 1.5\right) \ ,
\label{eq:result_RRab} \\
M_{\mathrm{RRc}} &= (0.004 \pm 0.021) \notag \\
&+ (-2.497 \pm 0.095)\cdot\left(\log P + 0.5\right) \notag \\
&+ (0.013 \pm 0.010)\cdot\left([\mathrm{Fe/H}] + 1.5\right) \, , \mathrm{and}
\label{eq:result_RRc} \\
M_{\mathrm{T2Cep}} &= (-3.407 \pm 0.027) \notag \\
&+ (-2.506 \pm 0.039)\cdot\left(\log P - 1\right) \notag \\
&+ (0.107 \pm 0.043)\cdot\left([\mathrm{Fe/H}] + 1.5\right) \ .
\label{eq:result_T2Cep}
\end{align}

The absolute magnitude of the fiducial candle ($\alpha_{0,\mathrm{RRab}}$) is calibrated to a precision of 0.020\,mag ($0.92\%$ in distance). RRc and T2Cep reach comparable accuracy, with total uncertainties of 0.021\,mag ($0.97\%$) and 0.027\,mag ($1.3\%$), respectively. These values match the precision of \gaia\ DR3-based calibrations for classical Cepheids \citep[$0.9\%$;][]{cruzreyes2023}. Among the three populations, RRc show the lowest intrinsic scatter, indicating that they may offer the highest precision under ideal conditions. However, the larger amplitudes and distinctive light curves of RRab make them more easily identifiable and therefore more practical for distance measurements.
 
The three PL slopes are broadly consistent across types, though not identical: RRab stars have a slope of $-2.582 \pm 0.047$\,mag, while RRc and T2Cep yield $-2.497 \pm 0.095$\,mag and $-2.506 \pm 0.039$\,mag, respectively. A small but statistically significant metallicity dependence is detected for RRab ($\alpha_M = -0.028 \pm 0.008$\,mag/dex), whereas RRc show a negligible effect ($\alpha_M = 0.013 \pm 0.010$\,mag/dex). In contrast, T2Cep display a more pronounced dependence of $0.107 \pm 0.043$\,mag/dex, albeit at lower significance.

These \baseline\ results represent the most accurate absolute calibrations to date for RRab, RRc, and T2Cep stars in the \gaia\ Wesenheit system. A direct comparison with the RRab calibration from \citet{garofalo2022} is presented in Sect.~\ref{sec:calibration_comparison}.

\subsubsection{Uncertainties and Markov Chain Monte Carlo} \label{sec:MCMC}
Because the problem is linear, the least-squares matrix formalism provides an adequate solution. Its main limitation is the requirement of a full covariance matrix, which in this case is simplified to a diagonal form $\mathbf{C}$ (Sect.~\ref{sec:methodolgy}). This assumption neglects potential correlations between data points (off-diagonal terms) but ensures analytical consistency under linearity and Gaussian-distributed errors. To test whether this simplification biases the results or underestimates the uncertainties, an MCMC analysis is carried out as an independent cross-check. The posterior distributions obtained with MCMC show excellent agreement with the \baseline\ least-squares solution. Their near-Gaussian shape and the close match in parameter uncertainties confirm that the model is well calibrated and that the error budget is robust. Together, these results demonstrate that the least-squares estimates provide a reliable and efficient characterization of the fit.

Posterior distributions for selected parameters are shown in Fig.~\ref{fig:MCMC}, together with direct comparisons to the analytical (least-squares) estimates. The figure displays the primary fit parameters for RRab and RRc stars (results for T2Cep are similar), as well as representative examples of distance parameters for one host and one anchor cluster. The agreement is excellent; for example, the absolute calibration of RRab in the \baseline\ fit is $\alpha_{0,\mathrm{RRab,LS}} = -0.300 \pm 0.020$\,mag from the least-squares method and $\alpha_{0,\mathrm{RRab,MCMC}} = -0.299 \pm 0.020$\,mag from the MCMC analysis. Across all parameters, differences between the two methods remain within $0.1\sigma$, confirming that the least-squares approach neither introduces significant bias nor underestimates uncertainties.

Most parameters show minimal mutual correlation. A strong anti-correlation is evident between the absolute calibration parameter $\alpha_{0,\mathrm{RRab}}$ and the distance-related terms $\mu$ and $\Delta\mu$, as expected: increasing the inferred distances implies intrinsically brighter stars, requiring lower $\alpha_{0,\mathrm{RRab}}$ values to remain consistent with the observed magnitudes. Mild correlations also appear between the period slopes (e.g., $\beta_\mathrm{RRab}$) and their associated metallicity terms (e.g., $\alpha_{M,\mathrm{RRab}}$), suggesting that these parameter pairs encode partially overlapping physical information. This motivates the inclusion of cross-terms such as $\beta_M\cdot[\textrm{Fe/H}]\cdot\log P$ in some model variants (see Sect.~\ref{sec:variants}).

Overall, the MCMC analysis confirms the robustness of the least-squares solution. The close agreement in both central values and uncertainties demonstrates that simplifying the covariance matrix to a diagonal form does not bias the results. The near-Gaussian posteriors indicate that the parameters are well constrained and free from strong degeneracies, while the consistency of the covariance estimates at the few-percent level provides confidence that the error budget is reliable. Together, these results confirm that the least-squares solution is unbiased, efficient, and statistically consistent with the full posterior distribution explored via MCMC.

\subsubsection{Homogeneous distances to 93 GCs} \label{sec:GC_distances}
\begin{figure*}[p]
    \centering
    \includegraphics[scale=0.92]{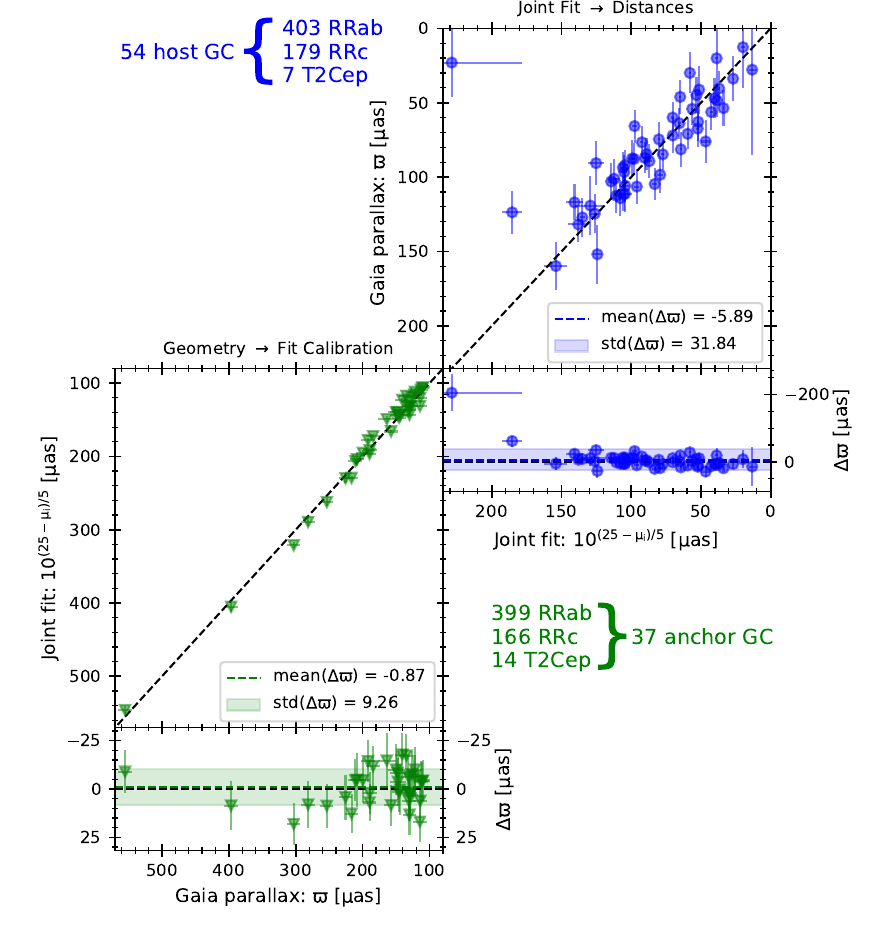}
    \caption{Visualization of the \baseline\ mini–distance ladder. The lower-left panel shows anchor GC parallaxes on the x-axis against the joint-fit distances on the y-axis, including residuals. The mean difference between \gaia\ parallax and fit parallax for anchor GCs is $-0.87 \pm 1.52\,\mu$as. The upper-right panel shows the host GC fit distances on the x-axis against the host GC parallaxes on the y-axis, with a mean difference of $-5.9 \pm 4.3\,\mu$as affected by outliers. The two outlier hosts Terzan~1 and BH~229 are due to sparse membership and do not affect the fit. The comparison reveals no evidence for a distance dependent bias of GDR3 parallaxes.}
    \label{fig:DL_copycat}
\end{figure*}
\begin{figure*}[p]
    \centering
    \includegraphics{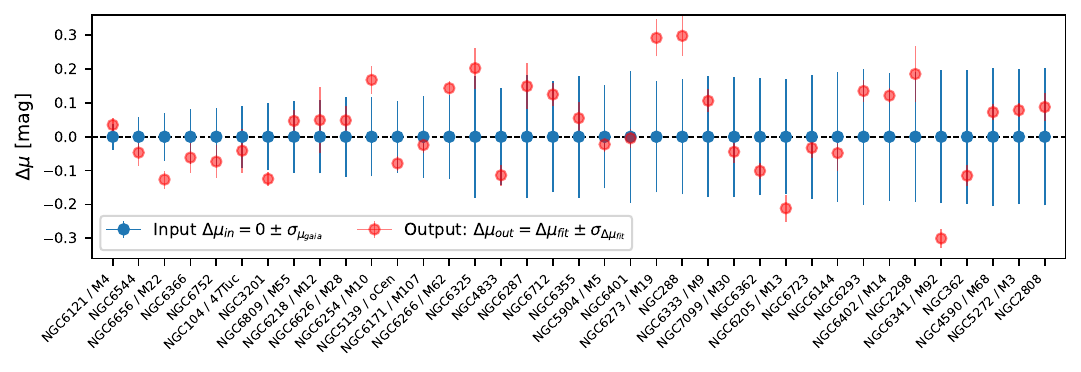}
    \caption{Correction parameters $\Delta\mu$ for each anchor GC in the \baseline. Blue symbols show the initial values (set to zero; Eq.~\ref{eq:constraint}), and red symbols the fitted values (Eq.~\ref{eq:q_fit}). Clusters are ordered by increasing distance, so the typical uncertainty in $\Delta\mu$ grows along the $x$-axis, although the number of member stars also contributes. Overall, 70\% of anchors remain within $1\sigma$ of their \gaia\ parallax values, and all lie within $2\sigma$.}
    \label{fig:delta_mu}
\end{figure*}

A key result of the \baseline\ analysis is a homogeneous determination of GC distances, calibrated to a zero point set by \gaia\ DR3 parallaxes. For anchors, the distances correspond to the geometric parallax-based estimates, modified by the fitted correction terms ($\Delta\mu$) from Eq.~\eqref{eq:constraint}. For hosts, the distance moduli ($\mu$) are inferred directly from the fit. Overall, \gaia\ parallaxes are consistent with the best-fit distances, aside from two identifiable outliers discussed below. Figure~\ref{fig:DL_copycat} illustrates the agreement between the \gaia-based and jointly fitted distances, highlighting their overall consistency.

For anchors, the methodology allows a direct comparison between fitted distances and \gaia-based values through the correction factor $\Delta\mu$ from Eq.~\eqref{eq:constraint}. This comparison assesses how well the fitted $\Delta\mu$ values remain consistent with zero, within the uncertainties of the \gaia\ parallax measurements for GC. 

This comparison is illustrated in Fig.~\ref{fig:delta_mu}, where both the initial and fitted values of $\Delta\mu$ are shown for each anchor, ordered by distance. As expected, the uncertainties increase with distance for both the \gaia\ parallaxes and the fitted correction parameters. Statistically, about 70\% of GC have correction factors within 1$\sigma$ of their \gaia\ values, and all anchors fall within 2$\sigma$ of their starting value.

A significant outlier appears in the case of NGC~6553, with a fitted $\Delta\mu$ offset nearly $6\sigma$ from zero. Closer inspection shows that this discrepancy arises from a single RRc candidate (\texttt{source\_id} = 4064796635702285056). Although the star is confirmed to be an RRc, its membership probabilities, both prior and posterior, indicate that it is unlikely to belong to NGC~6553. This star is therefore excluded from the final dataset, and since no other valid variables remain in that cluster, NGC~6553 is omitted from the final results. 

Overall, the distances derived from the fit closely match those obtained from \gaia\ parallaxes for the anchors. This agreement is corroborated in the bottom-left panel of Fig.~\ref{fig:DL_copycat}, where the mean parallax difference between \gaia\ measurements and the joint-fit values is $-0.87~\mu$as, with a standard deviation of $9.26~\mu$as. A similar consistency is observed for the hosts in the upper-right panel of the same figure, with a mean difference of $-5.89~\mu$as and a standard deviation of $31.84~\mu$as.

Caution is warranted when interpreting the comparison for the hosts, as the reliability of their \gaia\ parallaxes decreases at larger distances. Terzan~1 (HP2) and BH~229 (HP1), which appear as the two upper-left points in the plot, contain only 40 and 54 identified member stars, respectively. In such sparsely populated cases, membership assignments are less robust and may lead to biased cluster parallaxes. These GC are hosts rather than anchors, so their \gaia\ parallaxes are not used in the fit. Their uncertain parallaxes therefore do not affect the derived distances, which are determined from the fit itself. For example, the \gaia\ parallax for Terzan~1 is about $25~\mu$as, whereas both the fit and independent literature estimates give values in the $175$–$200~\mu$as range \citep{baumgardtvasiliev2021}.

\subsection{Variants} \label{sec:result_variants}
A comprehensive visual summary of the variant fits is provided in Appendix~\ref{app:variant_whiskers} (Fig.~\ref{fig:variant_whiskers}), while the numerical results are compiled in Table~\ref{tab:fit_results}. In total, 29 model variants, defined and categorized in Sect.~\ref{sec:variants}, are fitted to assess the impact of specific assumptions, data subsets, and potential sources of systematics.

In brief, all variants confirm the robustness of the \baseline\ solution. The fitted parameters remain consistent, with nearly all deviations lying within $1\sigma$ of the baseline values. This demonstrates that the absolute calibration is stable against changes in modeling assumptions or sample definitions. The whisker plots in the appendix further illustrate this point by showing both the statistical uncertainties of each variant and the narrow clustering of results around the baseline calibration.

\subsubsection{Outlier Rejection} \label{sec:variants_outlier_rejection}
The fit is remarkably robust once some level of outlier rejection is applied. The calibration stabilizes quickly for moderate thresholds and becomes effectively unchanged ($<1\sigma$ deviation) for stricter criteria. The specific value of the threshold $\kappa$ has only a minor effect on the final results, but the presence of an outlier-rejection step is essential. Omitting this step entirely introduces significant biases in several parameters; for instance, $\beta_\textrm{RRc}$ shifts by nearly $5\sigma$ relative to the \baseline\ model. In contrast, any variant that applies even a weak rejection (e.g., $\kappa = 5.0$) yields results that remain within $1\sigma$ of the baseline.

As expected, the choice of $\kappa$ has a pronounced impact on the goodness-of-fit metric. Without outlier rejection, the reduced chi-squared increases sharply to $\chi^2_\mathrm{dof} = 6.297$. A weak rejection threshold ($\kappa = 5.0$) reduces it to $\chi^2_\mathrm{dof} = 1.410$, while a strict cutoff ($\kappa = 2.5$) results in $\chi^2_\mathrm{dof} = 0.497$. These results confirm that the rejection step is critical to ensure a robust calibration, while the precise value of $\kappa$ has limited influence. Chauvenet’s criterion provides a principled, data-driven threshold that balances robustness with minimal distortion of the underlying distribution.

Figure~\ref{fig:variant_whiskers} reveals a subtle but systematic trend: the metallicity term $\alpha_{M,\mathrm{RRab}}$ decreases as the rejection threshold becomes stricter, which in turn yields slightly brighter absolute calibrations $\alpha_{0,\mathrm{RRab}}$. This behavior is expected, since both metallicity and outliers contribute to scatter in the PLR fit. Stricter rejection may inadvertently exclude stars with genuine residuals driven by metallicity effects. Overly aggressive rejection can therefore suppress intrinsic scatter associated with metallicity, artificially flattening $\alpha_M$ and biasing the calibration. These trends underscore the need for an objective, statistically motivated threshold. In this analysis, Chauvenet’s criterion provides such a compromise: it removes true outliers while minimizing the risk of discarding stars whose deviations arise from astrophysical causes such as metallicity variation. Moreover, a $5\sigma$ threshold yields results nearly identical to \baseline, whereas omitting rejection shifts the absolute calibration by $1\sigma$, indicating that the outliers affecting $\alpha_M$ lie beyond $5\sigma$ from the fit.

\subsubsection{Metallicities -- different sources for [Fe/H]}\label{sec:variants_metallicities}
Replacing the photometric metallicity calibration with that of \citet{muraveva2025} reduces the RRc sample by roughly half because the light-curve parameters $\phi_{31}$ and $A_2$ are frequently unavailable, especially in the hosts. This variant increases the uncertainty on $\alpha_M$ for RRc, although the results remain consistent within $1\sigma$.

Photometric metallicity estimates do not necessarily provide absolute values of [Fe/H], but rather proxies derived from light-curve morphology (via Fourier parameters). In practice, this means that the fit captures a correlation with light-curve shape rather than metallicity on an absolute scale. The absolute calibration of [Fe/H] is therefore not critical for determining GC distances. However, because these calibrations are light-curve dependent, any imperfections may induce a residual correlation between period and metallicity effects. 

In the variant using GC metallicities from \citet{harris2010}, the results again remain consistent with the \baseline. The uncertainties on the metallicity coefficients $\alpha_M$ increase significantly, by more than a factor of five for both RRab and RRc. This outcome is expected: GC metallicities are determined from red giant branch stars and may not capture variations among RRL. Assigning a single metallicity per cluster removes any intra-cluster variation that would otherwise help constrain $\alpha_M$, thereby reducing the leverage to measure metallicity dependence. In both cases, the absolute calibrations of the standard candles remain largely unchanged. 

Interestingly, the $\alpha_M$ term for T2Cep shifts by $\sim 2\sigma$ in this variant, even though the metallicity calibration was only altered for RRab and RRc. This behavior suggests that the metallicity dependence inferred for T2Cep is largely driven by the more numerous RRab, through their weight in the global fit. The change also reflects the rejection of a few additional T2Cep once cluster metallicities were adopted, further amplifying the effect. While the shift remains modest, it highlights that the metallicity dependence of T2Cep should be treated with caution.

\subsubsection{Parallaxes} \label{sec:variants_parallaxes}
The two variants that restrict the sample to narrower apparent-magnitude ranges ($G < 19$ and $14.5 < G < 18$) yield results nearly identical to those of the \baseline, despite visible inconsistencies between faint-star parallaxes and GC-level parallaxes (see Fig.~\ref{fig:Gaia_varpi_Gmag}, where $\langle\varpi_i - \langle \varpi\rangle_\mathrm{GC}\rangle$ reaches up to $-40\,\mu$as). This minimal impact is expected, as faint stars typically carry large uncertainties and therefore contribute little statistical weight to the GC parallax estimates.

By contrast, the variant that adopts parallaxes from VB21 produces a $1\sigma$ shift in the absolute calibration $\alpha_{0,\mathrm{RRab}}$. While both VB21 and the \baseline\ rely on \gaia\ parallaxes, they differ in the treatment of angular covariance and uncertainty propagation. As a result, the mean GC parallaxes used in VB21 are slightly offset from ours, with an average difference of $2~\mu$as across the anchor clusters.

This offset has a predictable effect: a change of $2~\mu$as in the input parallaxes translates into a $\sim$0.020\,mag shift in the absolute-magnitude calibration. This matches the observed difference in $\alpha_{0,\mathrm{RRab}}$ between the \baseline\ and the VB21 variant. A similar shift appears in the variant that uses median parallaxes instead of the mean.

\subsubsection{Models -- exploring different assumptions} \label{sec:variants_models}
Removing RRab reduces the $\chi^2_\mathrm{dof}$, while removing RRc increases it. This behavior arises from the use of a single extra-scatter term applied uniformly to all candles. As shown in Fig.~\ref{fig:baseline_residuals}, the residual spread is larger for RRab than for RRc. The shared extra-scatter value represents a compromise between the two (see Sect.~\ref{sec:LL_residuals}), which explains the observed changes in the reduced chi-square when either type is removed. While the extra-scatter term could be fine-tuned for each candle type to achieve $\chi^2_\mathrm{dof} \approx 1$ individually, it is kept fixed for simplicity. Overall, the model parameters demonstrate robustness, showing no significant changes when particular candles are excluded.

Model variants in which certain parameters are shared between standard candle types are also explored. Previous studies suggest that T2Cep may share their period slope with RRab \citep{braga2020}, whereas the present results indicate that the RRc slope is somewhat closer to that of T2Cep. Nevertheless, all three slopes remain consistent within $2\sigma$. To test these hypotheses, a variant enforcing $\beta_\mathrm{T2Cep} = \beta_\mathrm{RRab}$ is implemented. As expected, the fit converges to an intermediate value between the independently determined $\beta$ values from the \baseline, while the absolute calibrations of both types remain unchanged. Similar behavior is observed in variants with shared $\beta$ and $\alpha_M$ parameters, confirming that coupling these terms has minimal impact on the resulting absolute calibrations. The leverage on $\beta$ is, however, limited for RRab and RRc due to their narrow period ranges ($\log P \approx [-0.4, 0]$ for RRab and $\log P \approx [-0.6, -0.4]$ for RRc).

Several variants are used to assess the impact of metallicity on the fit results. The most direct approach involves removing the metallicity dependence entirely by setting $\alpha_M = 0$ for all standard candles. Under this constraint, the other parameters for RRab and RRc remain nearly unchanged, as expected, given that their fitted metallicity terms on the intercept ($\alpha_M$) are below 0.030\,mag/$\mathrm{dex}$. For T2Cep, however, $\alpha_M$ is more significant, and its removal results in a shift of $\beta_\mathrm{T2Cep}$ by 1.2$\sigma$. Despite these changes, the absolute calibrations of all candles remain stable, even when the metallicity dependence on the intercept is excluded.

Another set of variants introduces a metallicity dependence in the period slope by replacing $\beta$ with $\beta_0 + \beta_M[\textrm{Fe/H}]$ for both RRab and RRc, as has been suggested for classical Cepheid models \citep{anderson2016, khan2025}. This modification also allows probing the potential correlation observed in the MCMC results (see Fig.~\ref{fig:MCMC}) between $\beta$ and $\alpha_M$. For RRab, a significant metallicity dependence in the slope is detected, with $\beta_{0,2,\textrm{RRab}} = -0.393 \pm 0.078$, corresponding to a $5\sigma$ detection. In contrast, no significant effect is found for RRc, with $\beta_{0,2,\textrm{RRc}} = 0.048 \pm 0.110$. In this variant, the metallicity effect on the intercepts ($\alpha_M$) drops to zero, suggesting that the dependence initially attributed to $\alpha_M$ is effectively absorbed by the metallicity-dependent period slope term $\beta_M$. A separate variant where $\alpha_M$ is explicitly fixed to zero yields similar results. Repeating the analysis using GC metallicities from \citet{harris2010} shows that both $\beta_M$ terms are consistent with zero within $1\sigma$, indicating that the detection of metallicity dependence is sensitive to the adopted metallicity scale. This behavior most likely reflects limitations in the RRab photometric metallicity calibration of \citet{li2023}, underscoring the need for spectroscopic metallicities to reliably probe slope–metallicity dependencies. Nevertheless, in all of these variants, the absolute calibrations of the candles remain nearly unchanged.

\subsubsection{Pre-processing} \label{sec:variants_pre_processing}
In the variants where the extra-scatter term is either suppressed, halved, or doubled, the expected changes appear in the overall $\chi^2$, reflecting the adjusted uncertainties. These modifications, however, have only a marginal impact ($<1\sigma$) on the fitted parameters. Increasing the extra scatter primarily affects the uncertainties on the fitted parameters, particularly $\beta$ and $\alpha_M$, owing to the corresponding increase in $\chi^2$. These results support the choice of adopting a single value of $\sigma_\mathrm{extra} = 0.045\,\mathrm{mag}$.

\subsubsection{Anchors Criteria} \label{sec:variants_anchors_criteria}
Changing the anchor selection criteria alters the distribution of clusters classified as anchors versus hosts. Stricter thresholds may reclassify marginal cases, with clusters that previously met the anchor requirements instead treated as hosts, and vice versa, potentially affecting the global fit.

The extinction-based criterion has limited impact on the overall results. Removing it produces only minor changes to the anchor/host classification, and tightening the threshold to $E(B-V) = 0.5$ exerts negligible influence on the fitted parameters, despite a modest redistribution of clusters.

By contrast, the criterion based on signal-to-noise ratio ($S/N$) has a more significant effect. Doubling the $S/N$ threshold from 10 to 20 induces a shift of about $1.5\sigma$ in the absolute calibration, along with slightly increased parameter uncertainties. While these changes remain statistically acceptable, the stricter threshold substantially reduces the number of anchor clusters, thereby increasing the sensitivity of the fit to individual clusters. This reduced anchor sample can introduce systematics that would otherwise be averaged out in a broader distribution.

To balance sample size with calibration stability, the \baseline\ adopts an intermediate threshold of $S/N = 10$, which provides a compromise between selectivity and robustness. This value is also consistent with the typical limit above which \gaia\ parallaxes can be reliably inverted \citep{luri2018}.

\section{Discussion\label{sec:discussion}}
This section examines the residuals from the \baseline\ analysis (Sect.~\ref{sec:LL_residuals}). The RRab calibration is compared to similar results from the literature (Sect.~\ref{sec:calibration_comparison}), followed by a comparison of the derived GC distances with published values (Sect.~\ref{sec:comp_gc_distances}). Finally, the impact of metallicity and the possibility of metallicity inhomogeneities within clusters are discussed (Sect.~\ref{sec:metallicity_inhomogeneity}).

\subsection{Residuals} \label{sec:LL_residuals}
Period–luminosity relations exhibit intrinsic scatter due to the finite width of the instability strip \citep{marconi2015, bhardwaj2023}, which depends on wavelength \citep{madore2012,khan2025}. It is customary to introduce an extra-scatter term to account for this intrinsic scatter when fitting PL relations. For example, 0.07\,mag was adopted by \citet{riess2022} based on the LL residuals of LMC classical Cepheids observed in the infrared. However, \citet{riess2022gaia} reported even lower scatter among MW classical Cepheids in open clusters of $0.045\,$mag.

\begin{figure}
    \centering
    \includegraphics{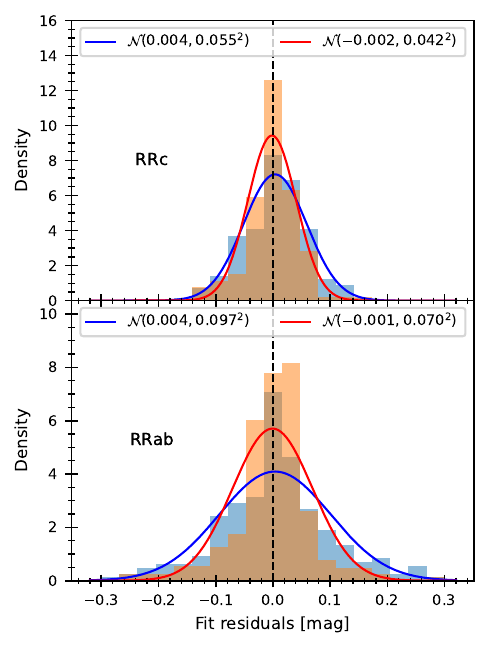}
    \caption{Residuals of the \baseline\ fit for RRab (bottom) and RRc (top). Blue histograms show host stars, orange histograms anchor stars. Overplotted are Gaussian distributions with identical mean and dispersion (blue for hosts, red for anchors). For anchor RRabs, a Kolmogorov–Smirnov test reveals significant deviations from Gaussianity, with a sharper central peak and extended tails. RRc residuals are both more Gaussian-like and systematically tighter than those of RRab.}

    \label{fig:baseline_residuals}
\end{figure}

Figure~\ref{fig:baseline_residuals} shows the distributions of fit residuals for RRab and RRc in anchor and host GCs. Most distributions are approximately Gaussian, except for RRab in anchors, which display a sharply peaked core and heavy tails, as confirmed by a Kolmogorov–Smirnov test. This deviation challenges the Gaussian error assumption. However, the MCMC analysis indicates that this did not bias the fit parameters.

We quantified the intrinsic scatter to apply using the inter-percentile range (16th–84th percentiles), which is more robust to non-Gaussianity than the standard deviation. For anchor RRab, this yielded $\sim$0.045\,mag, marginally larger than the scatter of RRc in anchors ($\sigma = 0.042$\,mag). For simplicity, we adopted a very small additional scatter of 0.045\,mag for all variables, as the number of T2Cep was insufficient for a meaningful estimate. 

Several physical and observational explanations for the extended tails of the RRab anchor residuals were considered. Depth effects within GCs are ruled out, since their physical sizes (a few parsecs) are negligible compared to their several kpc distances. The \citet{blazhko1907} effect, which induces amplitude and phase modulations in RRL light curves, was ruled out based on measured changes in mean magnitude over time \citep{skarka2020}. Crowding and blending, which can bias photometric measurements, would likely produce skewed residuals rather than the symmetric heavy-tailed structure observed, and would be more applicable in host than in anchor GCs. Correlations between pulsation amplitudes across the three GDR3 photometric bands were also examined but revealed no linkages with the shapes of the fit residuals. Thus far, none of the mechanisms explored adequately explain the observed deviations from Gaussianity.

Variants using different values of the extra-scatter term showed expected changes in overall $\chi^2$, while resulting in LL parameters well within $1\sigma$ of the \baseline. This robustness indicates that while the choice of scatter influences statistical measures of goodness-of-fit, it does not bias the core results. 

\subsection{Comparison with the literature} \label{sec:calibration_comparison}
The \baseline\ results can be compared with the RRL calibration reported by \citet{garofalo2022}, which was based on \gaia\ EDR3 parallaxes and \gaia\ DR2 photometry of field RRab: 
\begin{equation} 
    M_G^W = \left(-0.88^{+0.08}_{-0.09}\right) + \left(-2.49^{+0.21}_{-0.20}\right)\log P + \left(0.14^{+0.03}_{-0.03}\right)\mathrm{[\textrm{Fe/H}]} \ .
\end{equation}
A key difference with our results lies in the adopted pivot points: the present calibration is anchored at $\log P = -0.25$ and $[\textrm{Fe/H}] = -1.5$, whereas theirs is defined at $\log P = 0$ and solar metallicity. Shifting our \baseline\ fit to the same pivot values (Eq.~\ref{eq:host_RRab}, Table~\ref{tab:fit_results}) yields:
\begin{align}
    M_G^W &= \left(-0.986^{+0.034}_{-0.034}\right) + \left(-2.582^{+0.047}_{-0.047}\right)\log P \notag \\
    &+ \left(-0.028^{+0.008}_{-0.008}\right)\mathrm{[\textrm{Fe/H}]} \ .
\end{align}

Thus we found the two calibrations to be broadly consistent in both the absolute magnitude zero point ($\alpha'_{0,\mathrm{RRab}}$; agreement within $1.2\sigma$) and the period slope ($\beta_\mathrm{RRab}$; agreement within $1\sigma$). However, the uncertainties from the present analysis are substantially smaller.

The most significant difference lies in the metallicity term. The calibration reported by \citet{garofalo2022} implies a strong positive dependence of $\alpha_{M,\mathrm{RRab}} = 0.14 \pm 0.03$~mag/dex, corresponding to a $4.5\sigma$ detection. In contrast, the \baseline\ solution yields a negative dependence of $\alpha_{M,\mathrm{RRab}} = -0.028 \pm 0.008$~mag/dex, significant at the $3.5\sigma$ level and differing from their value by $5.4\sigma$. However, as discussed in Sect.~\ref{sec:variants_metallicities}, the present analysis indicates that metallicity plays only a minor role in the absolute calibration of the candles.

Several methodological differences may underlie this discrepancy. Firstly, the metallicities adopted in \citet{garofalo2022} were taken from \citet{muraveva2018}, who, in turn, had adopted values from \citet{dambis2013}. This latter catalog combined metallicity estimates from a variety of heterogeneous sources and methods and  homogenized them onto a common scale. By contrast, our \baseline\ analysis uses internally consistent photometric metallicities calibrated based on GDR3 light-curve parameters and homogeneous spectroscopic observations \citep{li2023}. Secondly, the calibration based on bright (\(G \lesssim 13\,\mathrm{mag}\)) field RRab in \citet{garofalo2022} was not corrected for residual parallax systematics, which have been investigated in detail but remain complex in several respects \citep[and references therein]{khan2023b}. Lastly, adopting mean-centered photometric metallicities minimizes parameter correlations and ensures that $\alpha_0$ corresponds most closely to the observed stellar sample. 

\subsection{Comparison of GC distances with the literature} \label{sec:comp_gc_distances}
\begin{figure}
    \centering
    \includegraphics{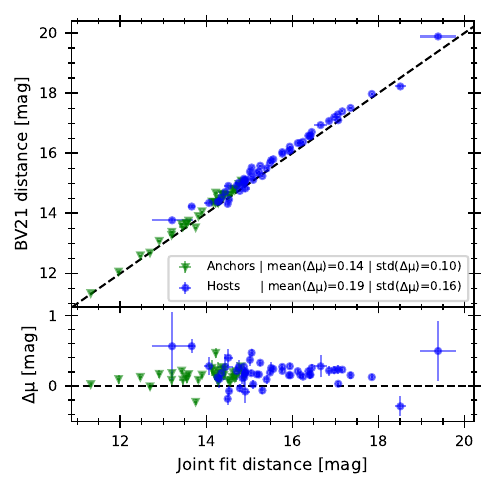}
    \caption{Comparison of distance moduli derived in this work for host (blue) and anchor (green) with literature values from \citet{baumgardtvasiliev2021}. The top panel shows the direct comparison, while the bottom panel presents the residuals.}
    \label{fig:literature_comparison_parallax}
\end{figure}

\citet[BV21]{baumgardtvasiliev2021} presented a catalog containing GC parameters, including distances featuring small uncertainties. The distances of these GCs were determined by averaging literature distance estimates from a variety of sources, often including highly correlated information, such as multiple estimates based on the same RRL member stars (e.g., analyzed in different studies). For example, the BV21 distance of NGC~5139 was estimated based on 27 published values, 14 of which were derived from RRL distances that share underlying assumptions and datasets. For each GC in the BV21 sample, the details of this averaging should be carefully considered as the available literature information varies from case to case. Compared to the GC distances in BV21, our homogeneously determined distances yield a mean difference of $\Delta\mu = 0.15$\,mag among anchors and $\Delta\mu = 0.17$\,mag among hosts, with dispersions of $0.08$ and $0.15$\,mag, respectively. Figure~\ref{fig:literature_comparison_parallax} illustrates this comparison.

Based on the GC distances in BV21, \citet{vasilievbaumgardt2021} raised concerns about a possible distance-dependent bias of \gaia\ parallaxes in the low-S/N regime. However, our joint fit to the 93 GCs in our sample does not confirm such a trend. Instead, our mini-distance ladder in Fig.\,\ref{fig:DL_copycat} shows no trend when comparing measured GC parallaxes to parallaxes obtained by inverting the precise luminosity distances by our fit. Thus, we find GDR3 GC parallaxes to remain reliable even at low S/N.

The RRL PLR calibrations presented by \citet{bhardwaj2023}, based on near-infrared photometry and anchored to field RRL, yield GC distances consistent with BV21. However, when applied to the LMC, these calibrations predict distances that are $0.06$–$0.12$\,mag larger than the detached eclipsing binary (DEB) distance of \citet{pietrzyski2019}. A similar offset was reported by \citet{sicignano2024}, who found $\Delta\mu \approx 0.09$–$0.15$\,mag between GC distances derived from LMC-calibrated T2Cep and those of BV21. Taken together, these results indicate that the DEB distance to the LMC is in tension with the BV21 GC distances inferred from both RRL and T2Cep.

The methodology adopted by \citet{bhardwaj2017} differs from ours in several key respects. Their analysis relied on distances from the \citet{bailerjones2021} catalog, which employs a Bayesian framework combining GDR3 parallaxes, Galactic priors, and bias corrections following \citet{lindegren2021}. Moreover, their sample is dominated by bright stars: 76\% of RRab and 87\% of RRc have $G<13$\,mag, a regime where residual parallax offsets are known to be significant \citep[see][and references therein]{khan2023b}. For a typical GC anchor at 5 kpc ($\varpi=200\,\mu$as), a systematic parallax shift of $+10\,\mu$as would translate into a distance modulus shift of $\approx-0.10$\,mag, thereby reconciling the GC distances with the LMC DEB distance. This suggests that the discrepancy between our results and those of \citet{bhardwaj2023} is plausibly driven by residual parallax bias.

In summary, the systematic differences between our homogeneously determined GC distances, anchored to the parallaxes of \citet{cruzreyes2024}, and those reported in the recent literature are best explained by (i) the averaging of heterogeneous and correlated values in the case of BV21, and (ii) uncorrected residual systematics in GDR3 parallaxes in the case of \citet{bhardwaj2023}.

\subsection{Metallicity effect and possible GC inhomogeneity} \label{sec:metallicity_inhomogeneity}
Our analysis reveals some limitations with respect to RRL metallicities. However, these limitations did not significantly affect the absolute calibrations of the standard candle intercepts.

Firstly, Fig.~\ref{fig:variant_whiskers} shows a trend between $\alpha_{0,\mathrm{RRab}}$, $\alpha_{M,\mathrm{RRab}}$, and the outlier-rejection threshold $\kappa$ (Sect.~\ref{sec:variants_outlier_rejection}). While we have adopted Chauvenet's as an objective criterion for outlier rejection, we cannot distinguish whether inaccurate photometric metallicities or chemical inhomogeneities create the excess residual scatter. The former possibility is corroborated by alternative models (Sect.~\ref{sec:variants_models}) involving a slope-metallicity dependence for RRab stars, which results in a putatively significant ($5\sigma$) detection of $\beta_{\mathrm{M,RRAb}} \ne 0$. Although a similar effect is predicted for classical Cepheids based on stellar evolution models \citep{khan2025}, we stress that we consider the $\beta_{\mathrm{M,RRAb}} \ne 0$ result spurious, as it appears only for RRab stars with photometric metallicities and disappears when GC metallicities are considered instead. A period-dependent systematic in the calibration of photometric RRL metallicities based on $P$, $R_{21}$, and $\phi_{31}$ by \citet{li2023} could potentially explain this effect and appears to be a reasonable possibility given the inherent correlation between light curve shape and pulsation period.    
However, it is also possible that chemical inhomogeneities within GCs lead to excess LL residuals. In favor of this possibility is the observation that many GCs exhibit a larger spread in photometric metallicities than the reported scatter of their calibration. Specifically, \citet{li2023} reported typical scatters of $0.24$\,dex (RRab) and $0.16$\,dex (RRc) for their photometric [Fe/H] calibration. In contrast, of the 20 GCs hosting $>10$ RRab, 5 exhibit $\sigma_{\mathrm{[Fe/H]}}>0.3$\,dex; among the 9 GCs with $>10$ RRc, 6 show $\sigma_{\mathrm{[Fe/H]}}>0.3$\,dex. Extreme cases include NGC~6715 (M~54), which exhibits a $0.44$\,dex spread in photometric [Fe/H] across 43 RRab, and NGC~6266 (M~62) with a dispersion of $0.42$\,dex across 24 RRc. These dispersions exceed the expected calibration scatter, suggesting that at least some GCs may host genuine internal metallicity spreads, as independently established for both M54 \citep[based on spectroscopic abundances]{M54Carretta2010} and M62 \citep[where a helium spread of $\Delta Y=0.08$ was inferred to explain horizontal branch morphology]{M62Milone2015}. Evidence for chemical inhomogeneity among RRL in GCs was presented by \citet{cruzreyes2024} who showed that different helium abundances ($\Delta Y \approx 0.06$) were required to reproduce the instability-strip edges of RRab and RRc stars. This creates an exciting link to the study of multiple populations in GCs \citep{bastian2018}. Helium and light-element variations would shift stars across the instability strip \citep[e.g.,][]{marconi2015,marconi2018}, altering pulsation amplitudes and effective temperatures at fixed $P$. Such effects would naturally increase LL scatter when abundances are either not determined individually (e.g., when GC-wide values for [Fe/H] are adopted) or if light curve shapes depended on elements not considered in the spectroscopy-based calibration, such as helium. 

The two above paragraphs thus lead to the interesting possibility that inhomogeneities in helium abundance might introduce a period-dependent bias among the photometric iron abundances reported by \citet{li2023}. At present, this evidence remains circumstantial. However, detailed spectroscopic investigations of RRL stars in GCs would allow to directly test for intra-cluster inhomogeneity.

\section{Summary and conclusions} \label{sec:conclusions}
This \emph{article} presents the first joint absolute calibration of the Leavitt Laws of RRab, RRc, and T2Cep stars. Our approach combines a linear least-squares framework, inspired by the SH0ES distance-ladder methodology, with an MCMC validation pipeline to assess uncertainties and robustness. The joint calibration of RRab, RRc, and T2Cep bring several advantages, notably through the ability of including a large number of GCs.

Our \baseline\ analysis simultaneously calibrates LLs of RRab, RRc, and T2Cep stars using reddening-free GDR3 Wesenheit magnitudes. The fiducial intercepts of RRab and RRc stars were thus calibrated to a distance accuracy of better than $1.0\%$, and T2Cep to $1.3\%$. We demonstrated robustness of this calibration using 31 fit variants exploring various systematics, including different parallax selections and metallicity treatments. A key result of our analysis are accurate and homogeneous distances to 93 GCs. These distances demonstrate that GC parallaxes based on GDR3 are not biased at the low-S/N end (i.e., as a function of distance). This is in contrast to previous claims in the literature, which we discuss in detail. 

Our results were less clear in terms of the metallicity dependencies of RRL and T2Cep. For RRL stars, we used photometric metallicities from \citet{li2023}. While it is clear that the effect is very small, it is barely significant ($3.5\sigma$) for RRab stars and not significant ($1.3\sigma$) for RRc stars. For T2Cep, the metallicity effect was considered based on spectroscopic [Fe/H] for each cluster, resulting in a $2\sigma$ signal. As discussed in Sect.\,\ref{sec:metallicity_inhomogeneity}, these issues do not substantially affect the calibration of our LL intercepts. However, we point out that further work is required to understand the metallicity dependencies of population-II pulsators, in particular in GCs where we find tentative signs of chemical inhomogeneities among RRL in individual clusters.

Our work shows the high potential of population-II pulsators for high accuracy distance determination. Future work will incorporate parallaxes of field RRL stars and measure the LMC's distance in order to strengthen the extragalactic distance scale by anchoring it to a large number of additional high-quality trigonometric parallaxes from \gaia. 

\section*{Data Availability}
The results presented in Tables~\ref{tab:result_anchor} and \ref{tab:result_host} will be made publicly available upon acceptance of this article.

\begin{acknowledgements}
This research has received support from the European Research Council (ERC) under the European Union's Horizon 2020 research and innovation programme (Grant Agreement No. 947660). RIA is funded by a Swiss National Science Foundation Eccellenza Professorial Fellowship (award PCEFP2\_194638).  

This work has made use of data from the European Space Agency (ESA) mission {\it Gaia} (\url{https://www.cosmos.esa.int/gaia}), processed by the {\it Gaia} Data Processing and Analysis Consortium (DPAC,
\url{https://www.cosmos.esa.int/web/gaia/dpac/consortium}). Funding for the DPAC has been provided by national institutions, in particular the institutions participating in the {\it Gaia} Multilateral Agreement. 

This research has made use of NASA's Astrophysics Data System; the SIMBAD database and the VizieR catalog access tool\footnote{\url{http://cdsweb.u-strasbg.fr/}} provided by CDS, Strasbourg.

This work made use of the following open-source software: 
\texttt{Python}\footnote{\url{https://www.python.org}}, 
\texttt{NumPy} \citep{harris2020array}, 
\texttt{Pandas} \citep{reback2020pandas}, 
\texttt{Matplotlib} \citep{hunter2007matplotlib}, 
\texttt{SciPy} \citep{virtanen2020scipy}, 
\texttt{Astropy} \citep{astropy:2013, astropy:2018, astropy:2022}, 
and \texttt{emcee} \citep{foreman-mackey2013emcee}. 
We are grateful to the developers and maintainers of these packages for making their work available to the community.
\end{acknowledgements}

\bibliographystyle{aa}
\bibliography{refs.bib}

\begin{appendix}
\section{Joint fit} \label{app:mathematical_details}
Equation \ref{eq:y=Lq} can be expressed in block form as this: 
\small
\begin{equation} \label{eq:y=Lq_full_form}
    \begin{bmatrix}
        \boldsymbol{m}^h_\mathrm{RRab}  \\
        (\boldsymbol{m}^a-\boldsymbol{\mu}_0)_\mathrm{RRab} \\
        \boldsymbol{m}^h_\mathrm{RRc} \\
        (\boldsymbol{m}^a-\boldsymbol{\mu}_0)_\mathrm{RRc} \\
        \boldsymbol{m}^h_\mathrm{T2C}  \\
        (\boldsymbol{m}^a-\boldsymbol{\mu}_0)_\mathrm{T2C}  \\
        \boldsymbol{0}
    \end{bmatrix}
    = 
    \begin{bmatrix}
        \mathds{I}^h_\mathrm{RRab} & 0 & \mathds{P}^h_\mathrm{RRab}  & 0 & 0 \\
        0 & \mathds{I}^a_\mathrm{RRab} & \mathds{P}^a_\mathrm{RRab}  & 0 & 0 \\
        \mathds{I}^h_\mathrm{RRc} & 0 & \mathds{B}^h_\mathrm{RRc} & \mathds{P}^h_\mathrm{RRc}  & 0 \\
        0 & \mathds{I}^a_\mathrm{RRc} & \mathds{B}^a_\mathrm{RRc}  & \mathds{P}^a_\mathrm{RRc}  & 0 \\
        \mathds{I}^h_\mathrm{T2C} & 0 & \mathds{B}^h_\mathrm{T2C}  & 0 & \mathds{P}^h_\mathrm{T2C}  \\
        0 & \mathds{I}^a_\mathrm{T2C} & \mathds{B}^a_\mathrm{T2C}  & 0 & \mathds{P}^a_\mathrm{T2C}  \\
        0 & \mathrm{I}\mathrm{d}& 0 & 0 & 0 \\
    \end{bmatrix}
    \cdot
    \begin{bmatrix}
        \boldsymbol{\mu} \\
        \boldsymbol{\Delta\mu} \\
        \boldsymbol{p}_\mathrm{RRab} \\
        \boldsymbol{p}_\mathrm{RRc} \\
        \boldsymbol{p}_\mathrm{T2C} \\
    \end{bmatrix}
\end{equation}
where bold symbols denote vectors (e.g., $\boldsymbol{m}^h$, $\boldsymbol{\mu}_0$), and blackboard bold letters represent matrices (e.g., $\mathds{I}^h$, $\mathds{P}^h$, $\mathrm{Id}$). The superscripts $h$ and $a$ refer to hosts and anchors, respectively, while subscripts distinguish the different types of variable stars.

In this representation, the model structure is straightforward to visualize. The first row of the matrix corresponds to the hosts for RRab as given by Eq.~\eqref{eq:host_RRab}, the second row to the anchors for RRab as described in Eq.~\eqref{eq:anchor_RRab}. Rows 3 and 4 follow the same logic for RRc (Eqs.~\eqref{eq:host_RRc/T2Cep} and \eqref{eq:anchor_RRc/T2Cep}), rows 5 and 6 for T2Cep (Eqs.~\eqref{eq:host_RRc/T2Cep} and \eqref{eq:anchor_RRc/T2Cep}), and the last row represents the constraints on the known geometric distances of the anchors as defined in Eq.~\eqref{eq:constraint}.

\subsection{Observation vector \texorpdfstring{$\boldsymbol{y}$}{y}}
The vector $\boldsymbol{y}$ consists of two components, $\boldsymbol{m}^h$ and $(\boldsymbol{m}^a-\boldsymbol{\mu}_0)$, repeated for each type of variable star, and a null vector $\boldsymbol{0}$. The component $\boldsymbol{m}^h$ contains the magnitudes of stars in all hosts and is further subdivided into sub-vectors $\boldsymbol{m}_i$ corresponding to each cluster $i$. Each sub-vector $\boldsymbol{m}_i$ contains the magnitudes $m_{i,j}$ of the stars $j$ within cluster $i$:
\begin{equation} \label{eq:m_full_form}
    \boldsymbol{m}^h
    = 
    \begin{bmatrix}
        \boldsymbol{m}^W_1 \\
        \vdots \\
        \boldsymbol{m}^W_n
    \end{bmatrix}_{\left(\sum_i^n n_i\times1\right)}
    \text{with }
    \boldsymbol{m}^W_i
    = 
    \begin{bmatrix}
        m^W_{i,1} \\
        \vdots \\
        m^W_{i,n_i}
    \end{bmatrix}_{\left(n_i\times1\right)}
\end{equation}
where $n$ is the total number of host GC and $n_i$ is the number of stars of a given type (RRab, RRc, or T2Cep) in host $i$.

The component $(\boldsymbol{m}^a-\boldsymbol{\mu}_0)$ represents the magnitudes of stars in all anchors, corrected by the known distance modulus $\mu_{0,i}$. Like the previous component, this vector is composed of sub-vectors $\boldsymbol{m}_i-\boldsymbol{\mu}_{0,i}$ for each anchor $i$, containing the corrected magnitudes $m_{i,j}-\mu_{0,i}$ for each star $j$ in anchor cluster $i$:
\begin{equation} \label{eq:m-mu_full_form}
    (\boldsymbol{m^a-\mu})
    = 
    \begin{bmatrix}
        \boldsymbol{m}^W_{1} - \boldsymbol{\mu}_{0,1} \\
        \vdots \\
        \boldsymbol{m}^W_{n'} - \boldsymbol{\mu}_{0,n'}
    \end{bmatrix}_{\left(\sum_i^{n'}n'_i\times1\right)}
    \text{with }
    \boldsymbol{m}^W_i
    = 
    \begin{bmatrix}
        m^W_{i,1} - \mu_{0,i} \\
        \vdots \\
        m^W_{i,n'_i} - \mu_{0,i}
    \end{bmatrix}_{\left(n'_i\times1\right)}
\end{equation}
where $n'$ is the total number of anchor GC and $n'_i$ the number of stars of a given type in anchor $i$. The last component of $\boldsymbol{y}$ is the null vector $\boldsymbol{0}$, with one entry for each anchor.

\subsection{Design matrix \texorpdfstring{$\boldsymbol{L}$}{L}}
The matrix $\mathds{L}$ is composed of two main components, $\mathds{I}$ and $\mathds{P}$, repeated for each type of variable star. The matrix $\mathds{I}^h$ is similar to an identity matrix but with its diagonal filled with vectors $\boldsymbol{1}$ (containing only ones), while the remaining elements are zeros:
\begin{equation} \label{eq:I_full_form}
    \mathds{I}^h =
    \begin{bmatrix}
        \boldsymbol{1}_{n_1} & \dots & \boldsymbol{0}_{n_1}  \\
        \vdots & \ddots & \vdots \\
        \boldsymbol{0}_{n_n}  & \dots & \boldsymbol{1}_{n_n} 
    \end{bmatrix}_{\left(\sum_i^{n}n_i\times n\right)}
    \text{with }
    \boldsymbol{1}_{n_i}  = 
    \begin{bmatrix}
        1 \\
        \vdots \\
        1
    \end{bmatrix}_{\left(n_i\times1\right)}
    \text{and }
    \boldsymbol{0} = 
    \begin{bmatrix}
        0 \\
        \vdots \\
        0
    \end{bmatrix}_{\left(n_i\times1\right)} \ .
\end{equation}

The matrix $\mathds{P}$ encodes information about properties (period and metallicity) for each host (or anchor) cluster. It consists of sub-vectors: $\boldsymbol{1}$ filled with ones, $\boldsymbol{P}_i$ containing the logarithms of periods $P_{i,j}$ relative to the pivot $P_0$, and $\left[\boldsymbol{\textrm{Fe/H}}\right]_i$ containing the metallicities relative to the pivot $\left[\textrm{Fe/H}\right]_0$:
\begin{align} 
    &&\mathds{P}^h &=
    \begin{bmatrix}
        \boldsymbol{1}_{n_1} & \boldsymbol{P}_1 & \left[\boldsymbol{\textrm{Fe/H}}\right]_1 \\
        \vdots & \vdots & \vdots \\
        \boldsymbol{1}_{n_n}  & \boldsymbol{P}_n  & \left[\boldsymbol{\textrm{Fe/H}}\right]_n
    \end{bmatrix}_{\left(\sum_i^{n}n_i\times 3\right)}
    \notag \\
    &\quad \text{with } &\boldsymbol{P}_i &=
    \begin{bmatrix}
        \log(P_{i,1})-\log(P_0) \\
        \vdots \\
        \log(P_{i,n_i})-\log(P_0) 
    \end{bmatrix}_{\left(n_i\times1\right)}
    \notag \\
    &\quad \text{and } &\left[\boldsymbol{\textrm{Fe/H}}\right]_i \quad &=
    \begin{bmatrix}
        \left[\textrm{Fe/H}\right]_{i,1}-\left[\textrm{Fe/H}\right]_0 \\
        \vdots \\
        \left[\textrm{Fe/H}\right]_{i,n_i}-\left[\textrm{Fe/H}\right]_0 
    \end{bmatrix}_{\left(n_i\times1\right)} \label{eq:P_full_form}
\end{align}

The matrix $\mathds{B}^h$ represents the absolute calibration of the primary candles ($\alpha_{0,\mathrm{RRab}}$) when propagated to other types of variables, and adopts a similar block structure:
\begin{align}
    &&\mathds{B}^h &=
    \begin{bmatrix}
        \boldsymbol{1}_{n_1} & 0 & 0 \\
        \vdots & \vdots & \vdots \\
        \boldsymbol{1}_{n_n}  & 0 & 0
    \end{bmatrix}_{\left(\sum_i^{n}n_i\times 3\right)}
    \notag \\
\end{align}

For all anchor components ($\mathds{I}^a$, $\mathds{P}^a$, $\mathds{B}^a$), the number of hosts $n$ and stars per host $n_i$ are replaced by the number of anchors $n'$ and stars per anchor $n'_i$. The final component, $\mathrm{Id}$, is an identity matrix of size $\left(n'\times n'\right)$, representing the geometric constraint.

\subsection{Parameter vector \texorpdfstring{$\boldsymbol{q}$}{q}}
The parameter vector $\boldsymbol{q}$ consists of three components: $\boldsymbol{\mu}$, $\boldsymbol{\Delta\mu}$, and $\boldsymbol{p}$, with the last one repeated for each type of variable star. The component $\boldsymbol{\mu}$ contains the fit parameters for the distances $\mu_i$ to the hosts, while $\boldsymbol{\Delta\mu}$ includes the correction parameters $\Delta\mu_i$ for the anchors. Finally, $\boldsymbol{p}$ contains the parameters describing the properties of a given type of variable star: the absolute magnitude $\alpha_0$ for a fiducial star at the pivot period and metallicity, the slope $\beta$ of the LL with respect to period, and the slope $\alpha_M$ of the PLR with respect to metallicity: 
\begin{equation}
    \boldsymbol{\mu}
    = 
    \begin{bmatrix}
        \mu_{1} \\
        \vdots \\
        \mu_{n}
    \end{bmatrix}_{\left(n\times1\right)}, 
    \quad
    \boldsymbol{\Delta\mu}
    = 
    \begin{bmatrix}
        \Delta\mu_{1'} \\
        \vdots \\
        \Delta\mu_{n'}
    \end{bmatrix}_{\left(n'\times1\right)},
    \quad
    \boldsymbol{p}
    = 
    \begin{bmatrix}
        \alpha_0 \textrm{ or }  \Delta\alpha_0\\
        \beta_0\\
        \alpha_M
    \end{bmatrix}_{\left(3\times1\right)}
\end{equation}

\subsection{Covariance Matrix \texorpdfstring{$\boldsymbol{C}$}{C}}
The covariance matrix $\boldsymbol{C}$ captures the uncertainties associated with the measurements in $\boldsymbol{y}$ and is approximated as diagonal. For the host and anchor stars, the uncertainties in $\boldsymbol{m}$ and $(\boldsymbol{m}-\boldsymbol{\mu})$ are given by the photometric uncertainties $\sigma(m^{W}_{i,j})$. For anchors, the geometric constraints are expressed as $\boldsymbol{0}$ in $\boldsymbol{y}$, with their uncertainties represented by $\sigma(\mu_{0,i})$ in the corresponding diagonal entries of $\boldsymbol{C}$:
\begin{equation} \label{eq:C}
    \boldsymbol{C} 
    = 
    \textrm{diag}
    \begin{bmatrix}
        \sigma\left(\boldsymbol{m}_{\mathrm{RRab}}^h\right) \\
        \sigma\left(\boldsymbol{m}_{\mathrm{RRab}}^a\right) \\
        \sigma\left(\boldsymbol{m}_{\mathrm{RRc}}^h\right) \\
        \sigma\left(\boldsymbol{m}_{\mathrm{RRc}}^a\right) \\
        \sigma\left(\boldsymbol{m}_{\mathrm{T2Cep}}^h\right) \\
        \sigma\left(\boldsymbol{m}_{\mathrm{T2Cep}}^a\right) \\
        \sigma\left(\boldsymbol{\mu}_0\right) 
    \end{bmatrix}
\end{equation}

This diagonal approximation neglects possible correlations between stars but greatly simplifies the fitting process. In Sect.~\ref{sec:MCMC}, we assess the validity of this assumption by comparing the baseline fit with an MCMC-based implementation.

\subsection{Least-Squares Solution and Parameter Estimation}
The system of equations in Eq.~\eqref{eq:y=Lq} is solved with a least-squares minimization. The goodness of fit is quantified with the $\chi^2$ statistic:
\begin{equation}\label{eq:solve_y=Lq}
    \chi^2=\left(\boldsymbol{y}-\boldsymbol{L}\boldsymbol{q}\right)^T\boldsymbol{C}^{-1}\left(\boldsymbol{y}-\boldsymbol{L}\boldsymbol{q}\right) \ .
\end{equation}

Minimization of $\chi^2$ yields the maximum-likelihood parameters:
\begin{equation}\label{eq:q_fit}
    \boldsymbol{q}_{\mathrm{best}} = \left(\boldsymbol{L}^T\boldsymbol{C}^{-1}\boldsymbol{L}\right)^{-1}\boldsymbol{L}^T\boldsymbol{C}^{-1}\boldsymbol{y} \ .
\end{equation}

The associated covariance matrix, which encodes the uncertainties of the fitted parameters, is
\begin{equation}\label{eq:Sigma_fit}
    \boldsymbol{\Sigma}_{\mathrm{best}} = \left(\boldsymbol{L}^T\boldsymbol{C}^{-1}\boldsymbol{L}\right)^{-1} \ .
\end{equation}

\section{Outlier rejection algorithm} \label{app:outlier_algorithm}
This appendix presents the detailed formulation of the outlier rejection algorithm used in Sect.~\ref{sec:outliers}. The procedure is summarized in Algorithm~\ref{alg:kappa}, which generalizes the classical $\sigma$-clipping approach to account for the population-dependent scatter of different types of pulsating stars.

\begin{algorithm}
    \caption{Multi-Population Adaptive $\sigma$-Clipping}\label{alg:kappa}
    \KwData{$\boldsymbol{y}$}
    \KwResult{$\boldsymbol{\hat{q}}$}
    $\boldsymbol{L},\boldsymbol{\hat{q}} \gets \text{Fit}(\boldsymbol{y})$\;
    \For{\textnormal{population}$_i$}{ 
        \Comment{For each population, the error, standard deviation and most deviant point is computed}
        $\left.\boldsymbol{\Delta y}\right|_i \gets \left.\left(\boldsymbol{y}-\boldsymbol{L}\boldsymbol{\hat{q}}\right)\right|_i$ \Comment*[r]{Restrict to pop i}
        $\sigma_i \gets \text{std}(\left.\boldsymbol{\Delta y}\right|_i)$\;
        $W_i \gets \text{max}(\left|\boldsymbol{\Delta y}\right|_i)$\;
    }

    \While{($W_1 > \kappa_1\cdot\sigma_1$) \textnormal{or} \dots \textnormal{or} ($W_n > \kappa_n\cdot\sigma_n$)}{
        \Comment{Iterate as long as any population has an outlier}
        $\boldsymbol{y} \gets $ Delete $y_i$ from $\boldsymbol{y}$, where $\Delta y_i$ is the most deviant point, if and only if ($W_i > \kappa_i \cdot \sigma_i$). Otherwise, check the second most deviant point, then the third, and so on, until a point satisfies the condition\;
        $\boldsymbol{L},\boldsymbol{\hat{q}} \gets \text{Fit}(\boldsymbol{y})$\;
        \For{\textnormal{population}$_i$}{ 
        $\left.\boldsymbol{\Delta y}\right|_i \gets \left.\left(\boldsymbol{y}-\boldsymbol{L}\boldsymbol{\hat{q}}\right)\right|_i$\;
        $\sigma_i \gets \text{std}(\left.\boldsymbol{\Delta y}\right|_i)$\;
        $W_i \gets \text{max}(\left|\boldsymbol{\Delta y}\right|_i)$\;
    }
      }
\end{algorithm}

\section{MCMC corner plot} \label{app:MCMC}
\begin{figure*}
    \centering
    \includegraphics{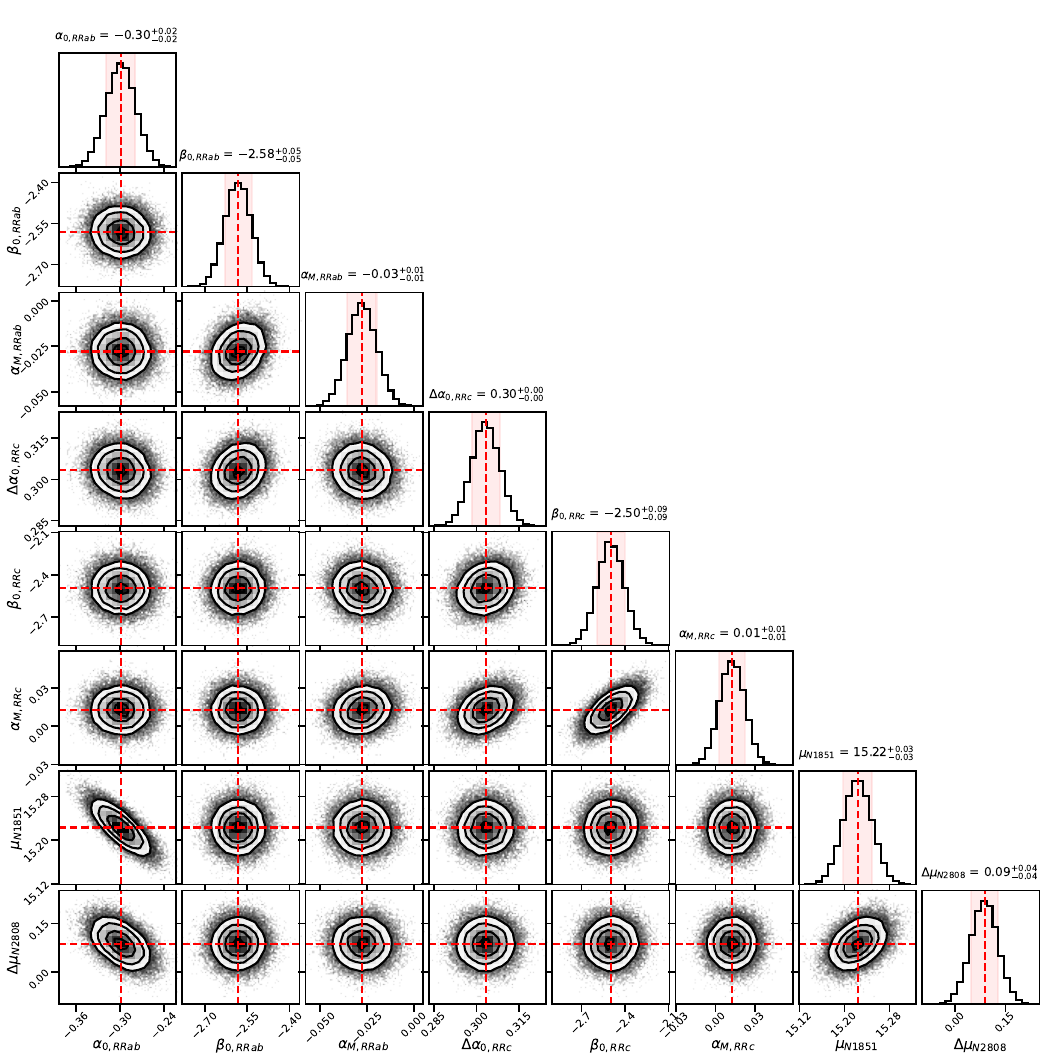}
    \caption{MCMC sampling of the $\chi^2$ statistic for the global fit, showing posterior distributions of selected parameters. Shown are the candle parameters $\alpha_0$, $\beta$, and $\alpha_M$ for RRab and RRc stars. The corresponding parameters for T2Cep exhibit similar distributions and are omitted here for clarity. The plot also includes one representative host distance and one anchor correction factor. Posterior distributions (black) are well approximated by Gaussians, with means and dispersions closely matching the least-squares solution from the \baseline\ (red). Contours indicate the $1\sigma$, $2\sigma$, and $3\sigma$ confidence regions. Correlations between period slopes and metallicity terms are discussed further in Sect.~\ref{sec:variants_metallicities}.}
    \label{fig:MCMC}
\end{figure*}

For completeness, we provide in Fig.~\ref{fig:MCMC} the corner plot of the MCMC sampling described in Sect.~\ref{sec:meth:MCMC}, with results discussed in Sect.~\ref{sec:MCMC}. The figure illustrates the posterior distributions of selected parameters and their correlations.  

\section{Variant summary plots} \label{app:variant_whiskers}
\begin{figure*}
    \centering
    \resizebox{\hsize}{!}{\includegraphics{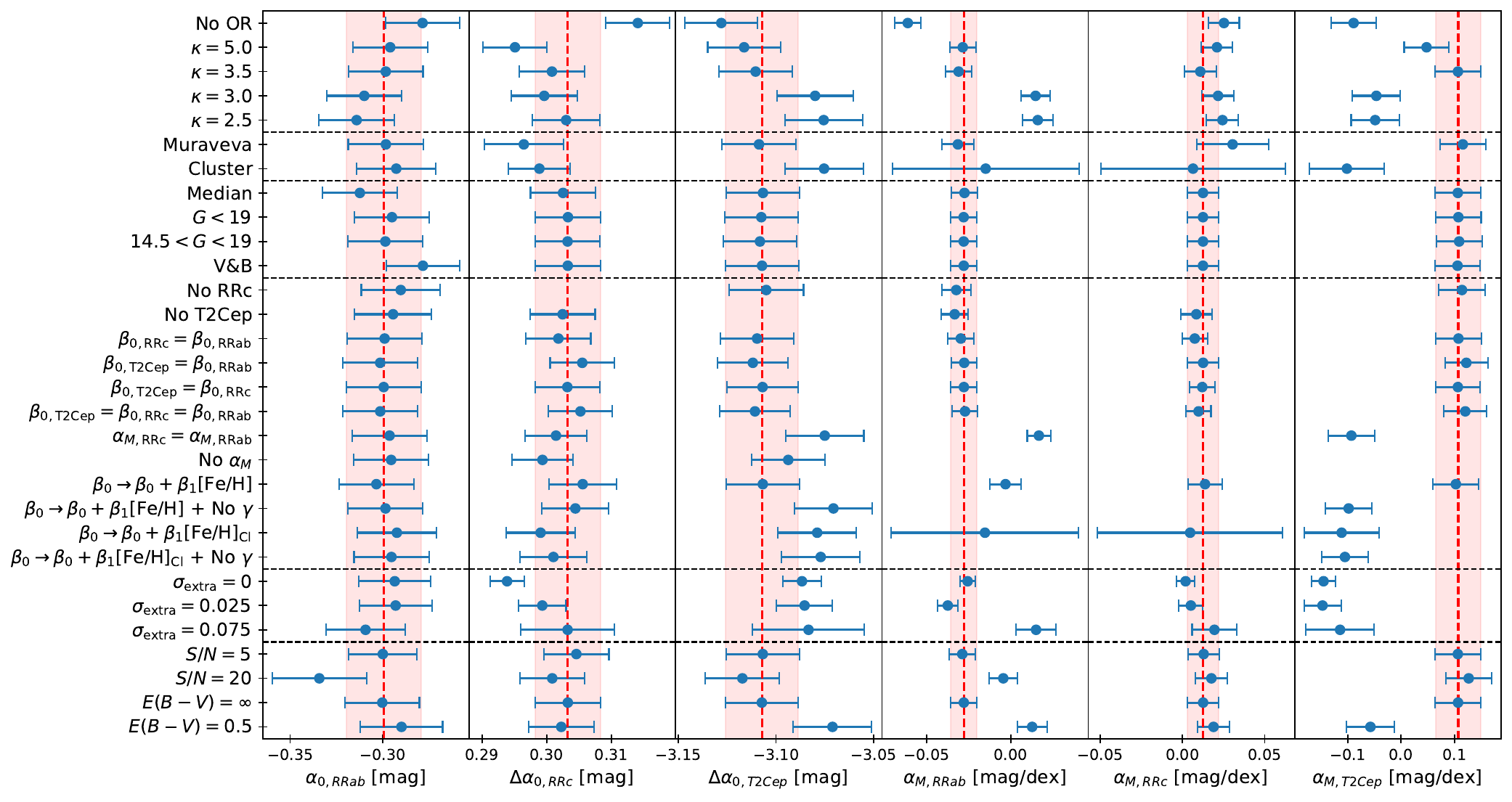}}
    \caption{Visual summary of the absolute calibration parameters ($\alpha_0$, $\Delta\alpha_0$) and metallicity terms ($\alpha_M$) for the standard candles across all model variants (excluding the 'No RRab' case, where the absolute scale and reference candle change). The red dashed line and shaded band indicate the baseline values and their associated uncertainties, while the blue markers and error bars represent the values obtained from each variant.}
    \label{fig:variant_whiskers}
\end{figure*}

Figure~\ref{fig:variant_whiskers} provides a visual summary of the absolute calibration parameters ($\alpha_0$, $\Delta\alpha_0$) and metallicity terms ($\alpha_M$) for the standard candles across all model variants (excluding the 'No RRab' case, where the absolute scale and reference candle change). 

\section{Result tables} \label{app:tables}
This appendix compiles the tables referenced throughout the analysis.  Table~\ref{tab:fit_results} summarizes the numerical results from all fits, including the \baseline\ solution and the variants described in Sect.~\ref{sec:variants}. Reported are the best-fit parameters for the different standard candles, together with the number of stars included in each fit and the corresponding reduced chi-squared values.  

Tables~\ref{tab:result_anchor} and~\ref{tab:result_host} list the GC classified as anchors and hosts, respectively. Each table provides the number of stars used, the distances derived from the \baseline\ fit, the mean parallaxes of the clusters, and the literature distances from \citet{baumgardtvasiliev2021}.

\clearpage
\begin{sidewaystable*}[ht]
\tiny
\renewcommand{\arraystretch}{1.2} readability
\setlength{\tabcolsep}{5.2pt}
\caption{Fitting results for each variant.} \label{tab:fit_results}
\begin{tabular}{|c|c|c|c c c c|c c c c|c c c c|}
\hline
\multirow{2}{*}{\textbf{Variant}} & \multirow{2}{*}{$\mathbf{\chi^2_{dof}}$} & \multirow{2}{*}{$\mathbf{N^a/N^h}$} & \multicolumn{4}{c|}{\textbf{RRab}} & \multicolumn{4}{c|}{\textbf{RRc}} & \multicolumn{4}{c|}{\textbf{T2Cep}} \\ \cline{4-15}
 &  &  & $N^a_*/N^h_*$ & $\alpha_0$ & $\beta_0$ & $\alpha_M$ & $N^a_*/N^h_*$ & $\Delta\alpha_0$ & $\beta_0$ & $\alpha_M$ & $N^a_*/N^h_*$ & $\Delta\alpha_0$ & $\beta_0$ & $\alpha_M$ \\ \hline
\textbf{Baseline} & \textbf{0.989} & \textbf{37/56} & \textbf{399/403} & \textbf{-0.300(20)} & \textbf{-2.582(47)} & \textbf{-0.028(8)} & \textbf{166/179} & \textbf{0.303(5)} & \textbf{-2.497(95)} & \textbf{0.013(10)} & \textbf{14/7} & \textbf{-3.107(19)} & \textbf{-2.506(39)} & \textbf{0.107(43)} \\ \hline
\textbf{MCMC}     & \textbf{0.989} & \textbf{37/56} & \textbf{399/403} & \textbf{-0.299(20)} & \textbf{-2.582(47)} & \textbf{-0.028(8)} & \textbf{166/179} & \textbf{0.303(5)} & \textbf{-2.497(94)} & \textbf{0.013(10)} & \textbf{14/7} & \textbf{-3.107(19)} & \textbf{-2.506(39)} & \textbf{0.107(42)} \\
\hline
\multicolumn{14}{l}{\textbf{Outlier Rejection}} \\ \hline
No OR & 6.297 & 37/56 & 407/437 & -0.278(20) & -2.497(47) & -0.061(8) & 172/197 & 0.314(5) & -2.052(94) & 0.025(9) & 14/7 & -3.128(19) & -2.292(38) & -0.089(42) \\
$\kappa=5.0$ & 1.410 & 37/56 & 404/424 & -0.296(20) & -2.574(47) & -0.028(8) & 170/192 & 0.295(5) & -2.387(94) & 0.021(10) & 14/7 & -3.116(19) & -2.494(39) & 0.048(42) \\
$\kappa=3.5$ & 1.069 & 37/56 & 400/405 & -0.298(20) & -2.581(47) & -0.031(8) & 166/185 & 0.301(5) & -2.499(95) & 0.011(10) & 14/7 & -3.110(19) & -2.510(39) & 0.107(43)\\
$\kappa=3.0$ & 0.723 & 37/56 & 392/387 & -0.310(20) & -2.540(48) & 0.015(9) & 165/175 & 0.300(5) & -2.509(96) & 0.022(10) & 14/7 & -3.080(19) & -2.421(40) & -0.046(45) \\
$\kappa=2.5$ & 0.497 & 36/56 & 366/335 & -0.314(20) & -2.560(49) & 0.016(9) & 160/166 & 0.303(5) & -2.561(97) & 0.024(10) & 12/7 & -3.076(20) & -2.431(41) & -0.049(45) \\ \hline
\multicolumn{14}{l}{\textbf{Metallicities}} \\ 
\hline
Muraveva & 1.047 & 37/53 & 399/402 & -0.298(20) & -2.599(50) & -0.031(10) & 92/74 & 0.296(6) & -2.302(138) & 0.030(22) & 14/7 & -3.109(19) & -2.495(40) & 0.116(43) \\ 
Cluster & 0.845 & 36/56 & 397/398 & -0.293(21) & -2.562(48) & -0.015(55) & 166/183 & 0.299(5) & -2.600(80) & 0.006(56) & 10/7 & -3.075(20) & -2.398(41) & -0.102(70) \\ \hline
\multicolumn{14}{l}{\textbf{Parallaxes}} \\ 
\hline
Median & 1.010 & 38/55 & 377/425 & -0.312(20) & -2.585(47) & -0.027(8) & 151/195 & 0.303(5) & -2.504(95) & 0.012(10) & 11/10 & -3.107(19) & -2.505(39) & 0.106(43) \\ 
$G<19$ & 0.990 & 37/56 & 399/403 & -0.295(20) & -2.581(47) & -0.028(8) & 166/179 & 0.303(5) & -2.497(95) & 0.012(10) & 14/7 & -3.108(19) & -2.506(39) & 0.107(43) \\
$14.5<G<19$  & 0.989 & 38/55 & 400/402 & -0.299(20) & -2.581(47) & -0.028(8) & 167/178 & 0.303(5) & -2.497(95) & 0.012(10) & 14/7 & -3.108(19) & -2.509(39) & 0.109(43) \\
VB21 & 0.989 & 40/53 & 402/400 & -0.278(20) & -2.580(47) & -0.028(8) & 168/177 & 0.303(5) & -2.498(95) & 0.012(10) & 14/7 & -3.107(19) & -2.503(39) & 0.106(43) \\
\hline
\multicolumn{14}{l}{\textbf{Models}} \\ 
\hline
No RRab\tablefootmark{*} & 0.578 & 29/43 & - & - & - & - & 164/182 & -0.002(23)\tablefootmark{*} & -2.526(102) & 0.013(11) & 7/7 & -3.365(25)\tablefootmark{*} & -2.328(67) & -0.198(70) \\
No RRc & 1.170 & 35/52 & 399/405 & -0.290(21) & -2.567(50) & -0.032(9) & - & - & - & - & 14/7 & -3.105(19) & -2.490(40) & 0.114(43) \\
No T2Cep & 0.897 & 34/55 & 399/402 & -0.294(21) & -2.564(47) & -0.033(8) & 166/182 & 0.302(5) & -2.508(95) & 0.008(10) & - & - & - & - \\ 
$\beta_{0,\textrm{RRc}}=\beta_{0,\textrm{RRab}}$ & 1.053 & 37/56 & 400/404 & -0.299(20) & -2.568(43) & -0.030(8) & 166/181 & 0.302(5) & RRab & 0.007(8) & 14/7 & -3.110(19) & -2.509(39) & 0.108(43) \\ 
$\beta_{0,\textrm{T2Cep}}=\beta_{0,\textrm{RRab}}$ & 1.018 & 37/56 & 399/404 & -0.301(20) & -2.534(30) & -0.028(8) & 166/179 & 0.305(5) & -2.487(95) & 0.013(10) & 14/7 & -3.112(18) & RRab & 0.122(40) \\
$\beta_{0,\textrm{T2Cep}}=\beta_{0,\textrm{RRc}}$ & 0.988 & 37/56 & 399/403 & -0.299(20) & -2.582(47) & -0.028(8) & 166/179 & 0.303(5) & -2.505(36) & 0.012(8) & 14/7 & -3.107(18) & RRc & 0.107(42) \\
$\beta_{0,\textrm{T2Cep}}=\beta_{0,\textrm{RRc}}=\beta_{0,\textrm{RRab}}$ & 1.017 & 37/56 & 399/404 & -0.301(20) & -2.530(29) & -0.027(8) & 166/179 & 0.305(5) & RRab & 0.010(8) & 14/7 & -3.111(18) & RRab & 0.120(40) \\
$\alpha_{M,\textrm{RRc}}=\alpha_{M,\textrm{RRab}}$ & 0.831 & 36/56 & 397/398 & -0.296(20) & -2.531(47) & 0.017(7) & 166/181 & 0.301(5) & -2.567(77) & RRab & 10/7 & -3.075(20) & -2.397(41) & -0.093(44) \\ 
No $\alpha_M$ & 0.857 & 36/56 & 398/403 & -0.295(20) & -2.554(47) & - & 166/181 & 0.299(5) & -2.647(71) & - & 10/7 & -3.094(19) & -2.454(34) & - \\ \hdashline
$\beta\rightarrow\beta_0 + \beta_M[\textrm{Fe/H}]$ & 0.964 & 37/56 & 399/402 & -0.303(20) & \makecell{-2.654(49)\\-0.393(78)} & -0.003(9) & 166/179 & 0.306(5) & \makecell{-2.505(96)\\0.048(110)} & 0.014(10) & 14/7 & -3.107(19) & -2.506(39) & 0.103(43) \\  \hdashline
\makecell{$\beta\rightarrow\beta_0 + \beta_M[\textrm{Fe/H}]$\\ + No $\alpha_M$} & 0.862 & 36/56 & 398/398 & -0.299(20) & \makecell{-2.590(50)\\-0.372(66)} & - & 166/183 & 0.304(5) & \makecell{-2.586(80)\\0.083(102)} & - & 10/7 & -3.071(20) & -2.401(41) & -0.098(44) \\ \hdashline
\makecell{$\beta\rightarrow\beta_0 + \beta_M[\textrm{Fe/H}]_{Cl}$} & 0.863 & 36/56 & 398/403 & -0.292(22) & \makecell{-2.565(49)\\0.015(151)} & -0.015(56) & 166/183 & 0.299(5) & \makecell{-2.599(81)\\0.059(171)} & 0.005(56) & 10/7 & -3.079(20) & -2.399(41) & -0.111(70) \\ \hdashline
\makecell{$\beta\rightarrow\beta_0 + \beta_M[\textrm{Fe/H}]_{Cl}$\\ + No $\alpha_M$} & 0.860 & 36/56 & 398/402 & -0.295(20) & \makecell{-2.546(47)\\-0.077(132)} & - & 166/183 & 0.301(5) & \makecell{-2.635(75)\\0.105(167)} & - & 10/7 & -3.077(20) & -2.400(41) & -0.105(44) \\ 
\hline
\multicolumn{14}{l}{\textbf{Pre-processing}} \\ \hline
$\sigma_{extra}=0$ & 36/56 & 2.283 & 400/401 & -0.293(19) & -2.538(26) & -0.026(4) & 166/181 & 0.294(3) & -2.495(50) & 0.002(6) & 10/6 & -3.087(10) & -2.418(22) & -0.145(23) \\
$\sigma_{extra}=0.025$ & 36/56 & 1.462 & 400/403 & -0.293(20) & -2.570(35) & -0.037(6) & 166/182 & 0.299(4) & -2.506(69) & 0.005(7) & 10/6 & -3.085(14) & -2.409(30) & -0.147(35) \\
$\sigma_{extra}=0.075$ & 36/56 & 0.482 & 396/401 & -0.309(21) & -2.524(68) & 0.015(12) & 166/181 & 0.303(7) & -2.551(138) & 0.019(14) & 12/7 & -3.083(29) & -2.394(57) & -0.114(64) \\
\hline
\multicolumn{14}{l}{\textbf{Anchors Criteria}} \\ \hline
$S/N=5$ & 1.024 & 71/22 & 608/195 & -0.300(19) & -2.570(47) & -0.029(8) & 269/76 & 0.305(5) & -2.494(95) & 0.013(10) & 18/3 & -3.107(19) & -2.508(39) & 0.106(43) \\
$S/N=20$ & 0.956 & 10/83 & 120/680 & -0.334(25) & -2.611(48) & -0.004(9) & 46/300 & 0.301(5) & -2.543(95) & 0.018(10) & 2/19 & -3.117(19) & -2.522(40) & 0.127(43) \\
$E(B-V)=\infty$ & 0.989 & 38/55 & 401/401 & -0.300(20) & -2.582(47) & -0.028(8) & 166/179 & 0.303(5) & -2.497(95) & 0.013(10) & 15/6 & -3.107(19) & -2.505(39) & 0.107(43) \\
$E(B-V)=0.5$ & 0.818 & 29/63 & 354/440 & -0.290(22) & -2.540(48) & 0.013(9) & 145/200 & 0.302(5) & -2.548(95) & 0.019(10) & 7/10 & -3.071(20) & -2.403(41) & -0.057(45) \\
\hline
\end{tabular}
\tablefoot{\tablefoottext{*}{In that case, the calibration of RRc is absolute (i.e., $\alpha_{0,\mathrm{RRc}}$) and not relative ($\Delta\alpha_{0,\mathrm{RRc}}$). Similarly, the calibration of T2Cep ($\Delta\alpha_{0,\mathrm{RRc}}$) is relative to RRc, as there is no RRab calibration.}}
\end{sidewaystable*}

\clearpage 
\begin{table*}
\centering
\caption{Anchor GCs used in the \baseline\ fit.\label{tab:result_anchor}} 
\begin{tabular}{@{}l|c|ccc|c|c|c|c}
\toprule
\multirow{2}{*}{\textbf{GC}} & \multirow{2}{*}{$\mathbf{N_{\star}}$} & \multicolumn{3}{c|}{\textbf{N}} & \multirow{2}{*}{$\boldsymbol{\mu}_\mathbf{fit}$ [mag]}&  \multirow{2}{*}{\textbf{$\boldsymbol{\mu}_\mathbf{BV21}$} [mag]} &
\multirow{2}{*}{$\boldsymbol{\pi}_\mathbf{fit}$ [$\mu$as]} &
\multirow{2}{*}{$\boldsymbol{\pi}_\mathbf{gaia}$ [$\mu$as]}\\ \cline{3-5}
& & \multicolumn{1}{c|}{\textbf{RRab}} & \multicolumn{1}{c|}{\textbf{RRc}} & \textbf{T2Cep} &  &  & &  \\ \hline
\midrule
NGC 104 (47Tuc) & 53656 & \multicolumn{1}{c|}{1} & \multicolumn{1}{c|}{0} & 0 & 13.19 $\pm$ 0.06 & 13.28 $\pm$ 0.01 & 229.97 $\pm$ 6.85 & 225.64 $\pm$ 9.53 \\
 NGC 288 & 5551 & \multicolumn{1}{c|}{1} & \multicolumn{1}{c|}{0} & 0 & 14.65 $\pm$ 0.06 & 14.77 $\pm$ 0.02 & 117.35 $\pm$ 3.18 & 134.62 $\pm$ 10.55 \\
 NGC 362 & 5887 & \multicolumn{1}{c|}{5} & \multicolumn{1}{c|}{1} & 0 & 14.60 $\pm$ 0.03 & 14.73 $\pm$ 0.02 & 119.99 $\pm$ 1.69 & 113.79 $\pm$ 10.37 \\
 NGC 2298 & 1370 & \multicolumn{1}{c|}{1} & \multicolumn{1}{c|}{0} & 0 & 14.76 $\pm$ 0.08 & 14.96 $\pm$ 0.04 & 111.91 $\pm$ 4.30 & 121.90 $\pm$ 10.75 \\
 \multicolumn{9}{c}{\ldots} \\
 \bottomrule
\end{tabular}
 
 \tablefoot{The table lists the number of stars associated with each cluster, the distance measured from the \baseline\ fit,the distance reported by \citet{baumgardtvasiliev2021} and the \gaia\ GC mean parallaxes (used in the fit for the anchors). The full table is available in electronic form via the CDS.}
\end{table*}
 
\begin{table*}
\centering
\caption{\label{tab:result_host}Host GCs used in the \baseline\ fit.} 
\begin{tabular}{@{}l|c|ccc|c|c|c|c}
\toprule
\multirow{2}{*}{\textbf{GC}} & \multirow{2}{*}{$\mathbf{N_{\star}}$} & \multicolumn{3}{c|}{\textbf{N}} & \multirow{2}{*}{$\boldsymbol{\mu}_\mathbf{fit}$ [mag]}&  \multirow{2}{*}{\textbf{$\boldsymbol{\mu}_\mathbf{BV21}$} [mag]} &
\multirow{2}{*}{$\boldsymbol{\pi}_\mathbf{fit}$ [$\mu$as]} &
\multirow{2}{*}{$\boldsymbol{\pi}_\mathbf{gaia}$ [$\mu$as]}\\  \cline{3-5}
 & & \multicolumn{1}{c|}{\textbf{RRab}} & \multicolumn{1}{c|}{\textbf{RRc}} & \textbf{T2Cep} &  &  &  \\ \hline
\midrule
NGC 1261 & 1892 & \multicolumn{1}{c|}{5} & \multicolumn{1}{c|}{3} & 0 & 15.92 $\pm$ 0.03 & 16.07 $\pm$ 0.03 & 65.44 $\pm$ 0.86 & 63.83 $\pm$ 10.61 \\
 NGC 1851 & 4201 & \multicolumn{1}{c|}{9} & \multicolumn{1}{c|}{4} & 0 & 15.22 $\pm$ 0.03 & 15.39 $\pm$ 0.02 & 90.22 $\pm$ 1.06 & 87.09 $\pm$ 10.41 \\
 NGC 1904 (M 79) & 2563 & \multicolumn{1}{c|}{2} & \multicolumn{1}{c|}{3} & 0 & 15.25 $\pm$ 0.03 & 15.58 $\pm$ 0.03 & 89.01 $\pm$ 1.34 & 84.53 $\pm$ 10.74 \\
 Pal 3 & 37 & \multicolumn{1}{c|}{1} & \multicolumn{1}{c|}{0} & 0 & 19.39 $\pm$ 0.42 & 19.88 $\pm$ 0.07 & 13.26 $\pm$ 2.55 & 27.78 $\pm$ 57.16 \\
 \multicolumn{9}{c}{\ldots}\\
 \bottomrule
 \end{tabular}
 \tablefoot{The table lists the number of stars associated with each cluster, the distance measured from the \baseline\ fit,the distance reported by \citet{baumgardtvasiliev2021} and the \gaia\ GC mean parallaxes (not used in the fit for the hosts). The full table is available in electronic form via the CDS.}
\end{table*}

\end{appendix}
\end{document}